\documentclass[preprint]{aastex}

\usepackage{hyperref}
\usepackage{epsfig}
\usepackage{natbib}
\usepackage{graphicx}                
\usepackage{lscape}
\usepackage{verbatim}              
\usepackage{amsmath,amsthm,amsfonts,amsopn,amssymb} 
\usepackage{longtable}
\usepackage{pdflscape}
\usepackage{afterpage}
\usepackage{rotating}					
\usepackage{ulem}
\usepackage{epstopdf}  
\usepackage{multirow}
\usepackage{color}
\usepackage[left]{lineno}
\received{}
\accepted{}

\slugcomment{to appear in ApJ}
\shorttitle{}
\shortauthors{Wang}

\begin{document}
\title{Influence of Stellar Multiplicity On Planet Formation. I. Evidence of Suppressed Planet Formation Due to Stellar Companions Within 20 AU and Validation of Four Planets From the $Kepler$ Multiple Planet Candidates}
\author{
Ji Wang\altaffilmark{1},
Ji-Wei Xie\altaffilmark{2,3},
Thomas Barclay\altaffilmark{4,5},
Debra A. Fischer\altaffilmark{1}
} 
\email{ji.wang@yale.edu}
%
\altaffiltext{1}{Department of Astronomy, Yale University, New Haven, CT 06511 USA}
\altaffiltext{2}{Department of Astronomy and Astrophysics, University of Toronto, Toronto, ON M5S 3H4, Canada}
\altaffiltext{3}{Department of Astronomy \& Key Laboratory of Modern Astronomy and Astrophysics in Ministry of Education, Nanjing University,
210093, China}
\altaffiltext{4}{NASA Ames Research Center, M/S 244-30, Moffett Field, CA 94035, USA}
\altaffiltext{5}{Bay Area Environmental Research Institute, Inc., 560 Third Street West, Sonoma, CA 95476, USA}

\begin{abstract}

The planet occurrence rate for multiple stars is important in two aspects. First, almost half of stellar systems in the solar neighborhood are multiple systems. Second, the comparison of the planet occurrence rate for multiple stars to that for single stars sheds light on the influence of stellar multiplicity on planet formation and evolution. We developed a method of distinguishing planet occurrence rates for single and multiple stars. From a sample of 138 bright (K$_P$$<$13.5) $Kepler$ multi-planet candidate systems, we compared the stellar multiplicity rate of these planet host stars to that of field stars. Using dynamical stability analyses and archival Doppler measurements, we find that the stellar multiplicity rate of planet host stars is significantly lower than field stars for semi-major axes less than 20 AU, suggesting that planet formation and evolution are suppressed by the presence of a close-in companion star at these separations. The influence of stellar multiplicity at larger separations is uncertain because of search incompleteness due to a limited Doppler observation time baseline and a lack of high resolution imaging observation. We calculated the planet confidence for the sample of mutlti-planet candidates, and find that the planet confidences for KOI 82.01, KOI  115.01, KOI 282.01 and KOI 1781.02 are higher than 99.7\% and thus validate the planetary nature of these four planet candidates. This sample of bright $Kepler$ multi-planet candidates with refined stellar and orbital parameters, planet confidence estimation, and nearby stellar companion identification offers a well-characterized sample for future theoretical and observational study. 

\end{abstract}


\section{Introduction}

The occurrence rate of planets in multiple stellar systems is an important factor when calculating the overall planet occurrence rate because almost half of the stellar systems in the solar neighborhood are multiple systems~\citep{Duquennoy1991, Fischer1992, Raghavan2010}. By comparing the planet occurrence rate of multiple stars and that of single stars, the influence of stellar multiplicity on planet formation, migration and evolution can be understood. In order to study the planet occurrence rate of multiple stars, there are two approaches: a dedicated survey for planets in multiple stellar systems can be carried out; or, the stellar multiplicity rate of stars with known planets can be determined and compared to a control sample. Adopting the first approach, surveys for planets in spectroscopic binaries were launched~\citep{Eggenberger2003, Eggenberger2007, Konacki2005, Konacki2009}. Using the second approach, the stellar multiplicity rate has been studied with imaging techniques, mostly with the Lucky Imaging and the adaptive optic (AO) techniques. 

Table \ref{tab:ReferencesList} summarizes previous works in the latter approach. 
The stellar multiplicity rate for planet host stars is less than or comparable to that of field stars in the solar neighborhood from previous works, but there are still a few issues that need to be addressed. Most of previous work focused on giant planets due to planet detection limit and the sample size was small. Studies of planets detected by the radial velocity (RV) technique suffer from the selection bias that close-in binaries are avoided in Doppler planet surveys. In addition, planet formation and evolution is most likely to be affected when stars have small separations~\citep{Desidera2007, Bonavita2007}, but stellar companions at small separations pose challenges for imaging techniques. The Doppler technique measuring stellar RV may be a more effective way of searching for stellar companions with small separations. 

The $Kepler$ mission has revolutionized the search for exoplanets and allows us to overcome several limitations. Since its launch in 2009, more than 5000 planet candidates have been discovered\footnote{http://exoplanetarchive.ipac.caltech.edu/}~\citep{Borucki2010,Borucki2011,Batalha2013,Burke2013}. Since the target selection process is not strongly discriminative to close-in binaries~\citep{Brown2011}, several circumbinary planets have been detected around binary stars with periods ranging from $\sim$10-40 days~\citep[e.g., ][]{Doyle2011,Orosz2012,Schwamb2013}. Therefore, the $Kepler$ mission provides us with a large sample of planet candidates without the severe selection bias against binary stars. 

Since the $Kepler$ era, planet occurrence rates have been addressed by many groups~\citep{Catanzarite2011, Youdin2011, Traub2012, Howard2012, Mann2012, Fressin2013, Dressing2013, Swift2013, Petigura2013, Gaidos2013, Kopparapu2013, Morton2013, Dong2013}. 
However, these works focused on an overall planet occurrence rate without distinguishing single stars and multiple stars. These two populations account for 54\% (single stars) and 46\% (multiple stars) of all stellar systems according to ~\citet{Raghavan2010}. Comparing planet occurrence rates for these two populations sheds light on how stellar multiplicity affects planet formation and evolution.

In this paper, we will compare the stellar multiplicity rate for a sample of bright stars with multi-planet candidates to the stellar multiplicity rate for field stars. We will also provide a method of distinguishing between planet occurrence rates for single and multiple stars. In \S \ref{sec:Method}, we will discuss the methods and analyses that are necessary for calculating the stellar multiplicity rate and estimating the planet confidence. We present an algorithm for obtaining a self consistent solution for stellar and orbital parameters that iterates between stellar evolution model and transiting light curve measurement in \S \ref{sec:iterative}. We conduct a search for stellar companions around planet host stars using the imaging techique in \S \ref{sec:ukirt}. We calculate the likelihood of a planet candidate being a bona fide planet in \S \ref{sec:confidence}. The calculation leads to the validation of 4 planet candidates with planet confidence higher than 99.7\%. We conduct a search for stellar companions using the Doppler technique based on archival data in \S \ref{sec:cfop}. We present a dynamical analysis to constrain possible stellar companions based on co-planarity of transiting multi-planet systems in \S \ref{sec:dynamical}. The results of the stellar multiplicity rate calculation and the planet occurrence rate calculation are given in \S \ref{sec:planet_frequency}. Discussion and summary will be given in \S \ref{sec:Summary}.

\section{Method and Analysis}
\label{sec:Method}

\subsection{Target Sample: Bright Multiple Planet Candidates From $Kepler$}
\label{sec:sample}

Our sample consists of bright host stars with multi-planet transiting systems from $Kepler$. Bright stars provide a higher signal to noise ratio (S/N) for photometric measurements, which helps to better determine stellar and orbital properties. Follow-up observations will be relatively easier for brighter stars. Multi-planet systems have lower false positive rates; they are more likely to be bona fide exoplanet systems. The presence of a second planet transit signal increases the likelihood of a bona fide planet by a factor of at least 25, and additional planet transit signals around one star provide a even larger boost of this likelihood~\citep{Lissauer2012}. Out of 5779 $Kepler$ Objects of Interest (KOIs)\footnote{http://exoplanetarchive.ipac.caltech.edu/}~\citep{Borucki2010, Batalha2013}, we selected all the systems with a $Kepler$ magnitude (K$_P$) brighter than 13.5 mag and with at least two planet candidates (not with false positive disposition). Most of these are FGK stars with 3700 K $\leq T_{\rm{eff}}\leq$ 7500 K and log $g$ higher than 4.0. The sample includes 343 planet candidates in 138 systems. For this sample, we have refined their stellar and orbital solutions, searched for stellar companions, and calculated the planet confidence levels. These analyses allow us to compare the stellar multiplicity rate with field stars and infer the planet occurrence rate for single and multiple star systems. 

\subsection{An Iterative Algorithm For Stellar and Orbital Parameters}
\label{sec:iterative}

\begin{figure}
\begin{center}
\includegraphics[width=16cm,height=8cm]{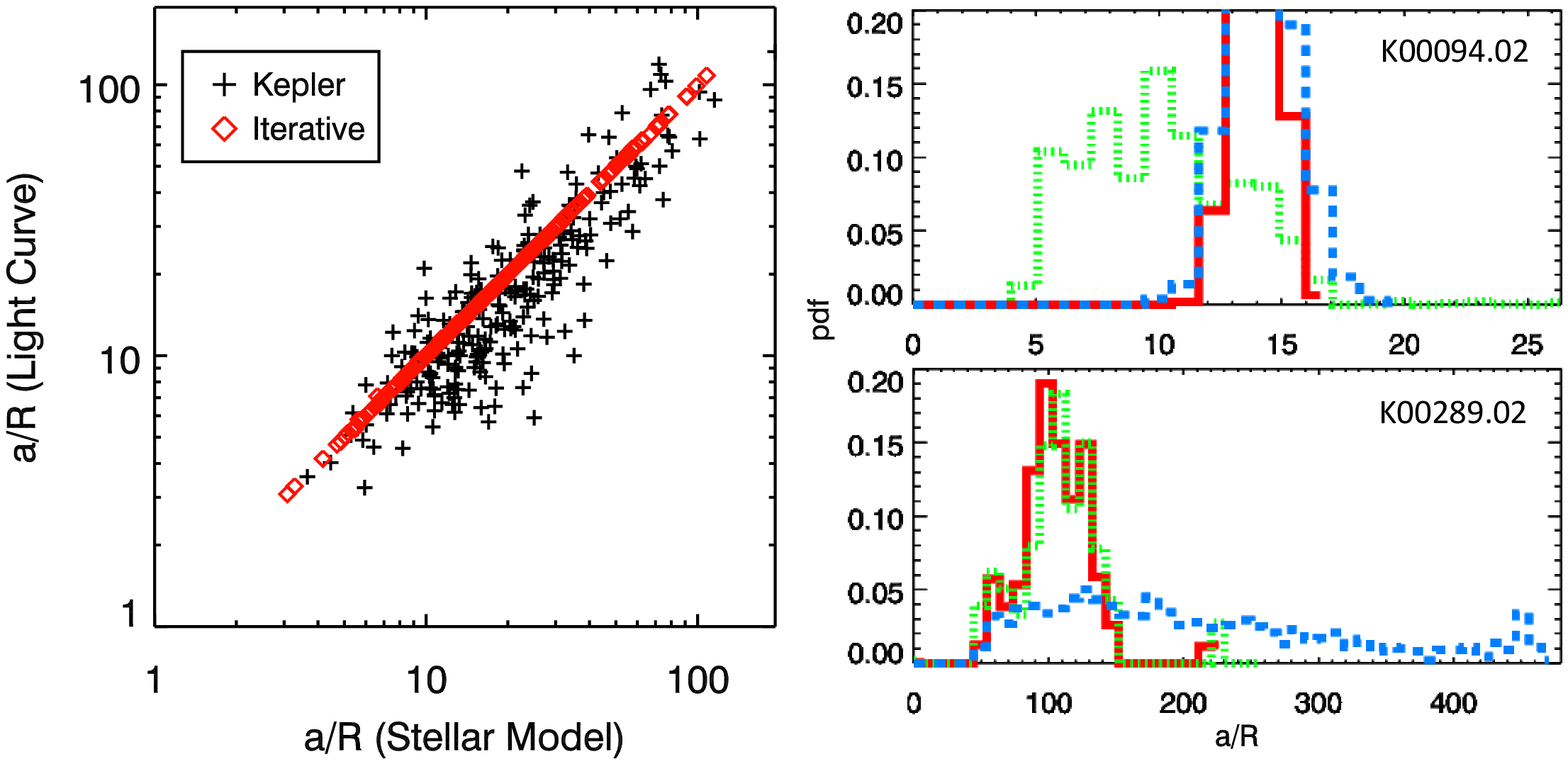} \caption{Left: comparison of $a/R_S$ between light curve derived values and stellar evolution model derived values. Black crosses represent $Kepler$ data and red diamonds represent the results of the iterative algorithm. The iterative algorithm will converge the $a/R_S$ from two approaches: light curve fitting and stellar evolution modeling, and reach a self-consistent solution of stellar and orbital properties. Top right: example of KOI 94.02, blue dashed line is the $a/R_S$ distribution from stellar evolution modeling, green dotted line is the $a/R_S$ distribution from the light curve fitting, red solid line is the converged distribution using the iterative algorithm. Bottom right is the same as top right except it is for KOI 289.02. 
\label{fig:Kepler_Iterative}}
\end{center}
\end{figure}

Stellar and orbital parameters from the KOI table are obtained independently from stellar evolution modeling and transit light curve analysis~\citep{Batalha2013}. Occasionally the calculations of some parameters are not consistent between stellar evolution modeling and transit light curve analysis. For example, $a/R_S$, the ratio of the planet orbital semi-major axis and the stellar radius, is sometimes different from these two analysis techniques. $a/R_S$ can be directly obtained from the transit light curve analysis based on a model described in ~\citet{Mandel2002}. We note that in their paper the notation $d/R_S$, the ratio of star-planet separation during transit and the stellar radius, is used and only circular orbits are considered. We used $a/R_S$ instead of $d/R_S$ and considered eccentric orbits in this paper. Alternatively, $a/R_S$ can be calculated from the Kepler's Third Law when the stellar mass and radius are known from a stellar evolution model and the orbital period is known from transit observations. Fig. \ref{fig:Kepler_Iterative} (left panel) shows the comparison of $a/R_S$ from these two calculations based on KOI values. The values for $a/R_S$ are not always consistent with each other, and can be off by a factor of 3-4, which is much larger than the predicted error bar. An algorithm that iterates between stellar evolution modeling and transit light curve analysis is needed to alleviate this discrepancy. 

We therefore developed an iterative algorithm that uses $a/R_S$ to link the results from the stellar evolution modeling and transit light curve analysis. A similar approach can also be found in~\citet{Torres2012,Dawson2012}. For stellar parameters, we used the Yonsei-Yale interpreter~\citep{Demarque2004} to calculate stellar mass, radius and luminosity based on the observation results reported in KOI table, such as effective temperature ($T_{\rm{eff}}$), surface gravity (log $g$), and metallicity ([Fe/H]). In a Monte-Carlo simulation, we generated a large sample of inputs (including $T_{\rm{eff}}$, log $g$, [Fe/H], $\alpha$ elements abundance, and age) for the Yonsei-Yale interpreter. The values for $T_{\rm{eff}}$, log $g$ and [Fe/H] follow a Gaussian distribution centered at the reported value from the KOI table and with a standard deviation equal to the reported error bar. For  $\alpha$ elements abundance and age, we assumed uniform distribution with a range of [0.0 dex, 0.2 dex] and [0.08 Gyr, 15 Gyr], respectively. We estimated the median values and the 68\% credible interval for stellar mass, radius, and luminosity based on their distributions from the outputs of the interpreter. With the orbital period, which is usually the most precisely determined parameter in a transit observation, the value and uncertainty of $a/R_S$ are calculated based on stellar evolution modeling. 

Once the distribution of $a/R_S$ is obtained from the stellar evolution modeling, it can be used as a prior for the estimation of orbital parameters in the transit light curve analysis. Instead of searching for $a/R_S$ values based purely on the light curves fitting, we considered only $a/R_S$ values that are consistent with the prior distribution. In the light curve fitting process, we used the model described in ~\citet{Mandel2002}. We adopted a boot-strapping algorithm to estimate the uncertainties of orbital parameters, in which we ran a larger number of trials in analyzing light curves that are perturbed based on the reported photometric measurement uncertainty. To avoid the dependence of results on the initial guess, we also perturbed the initial guess based on distributions of orbital parameters from previous runs. Initial guesses within 5-$\sigma$ dispersion were tried in order to avoid the sensitivity of the initial guess, and to explore a large parameter space. Orbital parameters are determined based on the distribution of the fitting results. The best fit values and their 68\% credible intervals are reported. A distribution of $a/R_S$ will be given at the end of the transit light curve analysis. The $a/R_S$ distribution will be fed into the Yonsei-Yale interpreter as a sampler. Only the outputs of the interpreter will be selected for $a/R_S$ distribution for the next round of iteration.

Fig. \ref{fig:Kepler_Iterative} (right panel) shows two examples of an $a/R_S$ distribution from stellar evolution modeling (blue dashed line) and transit light curve analysis (green dotted line) and the converged $a/R_S$ distribution from the iterative algorithm (red solid line). For KOI 94.02, because the stellar parameters (such as $T_{\rm{eff}}$, log $g$, and [Fe/H]) are determined by the SME analysis~\citep{Valenti2005}, the uncertainties for these parameters are smaller. Therefore, the constraint on $a/R_S$ from the stellar evolution modeling is tighter than that from the transit light curve analysis. Thus, the converged $a/R_S$ distribution is mainly determined by the $a/R_S$ distribution from the stellar evolution modeling. Conversely, for KOI 289.02, since $T_{\rm{eff}}$, log $g$, and [Fe/H] are determined by $J$-$K$ color, the constraint on $a/R_S$ from the stellar evolution modeling is weak, and the converged distribution of $a/R_S$ is mainly determined by the transiting light curve analysis.

\begin{figure}
\begin{center}
\includegraphics[width=16cm,height=16cm]{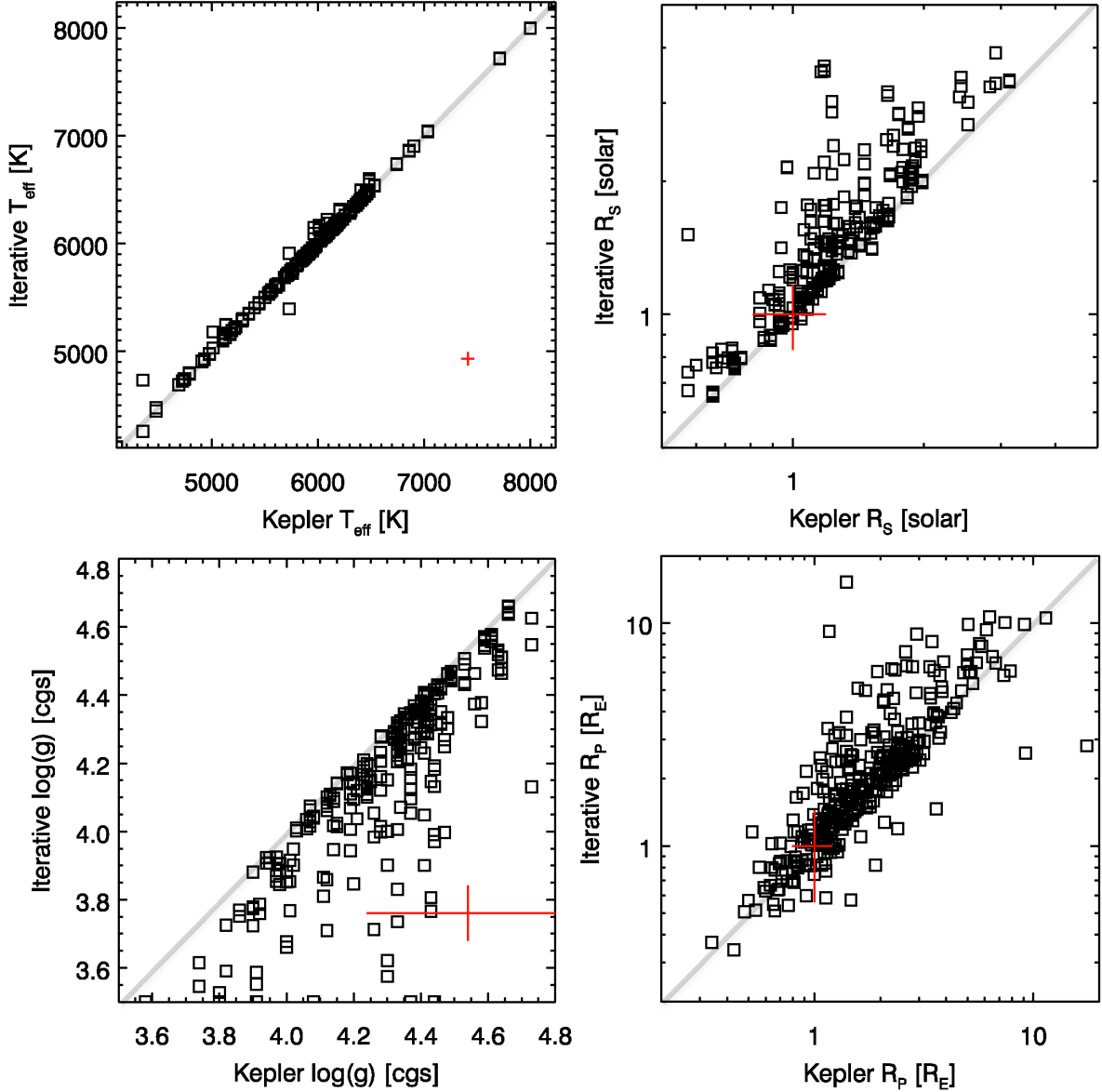} 
\caption{Comparison of $T_{\rm{eff}}$, $R_S$, $R_P$ and log(g) (clockwise from top left). Median error bar for each sub-plot is indicated by a red cross. The iterative algorithm results agree well with the KOI values for $T_{\rm{eff}}$. The comparison of $R_S$ and log(g) suggests that some of the KOI values for these two parameters  are overestimated (log(g)) or underestimated ($R_S$).
\label{fig:Kepler_Iterative_comp}}
\end{center}
\end{figure}

\begin{figure}
\begin{center}
\includegraphics[width=16cm,height=10cm]{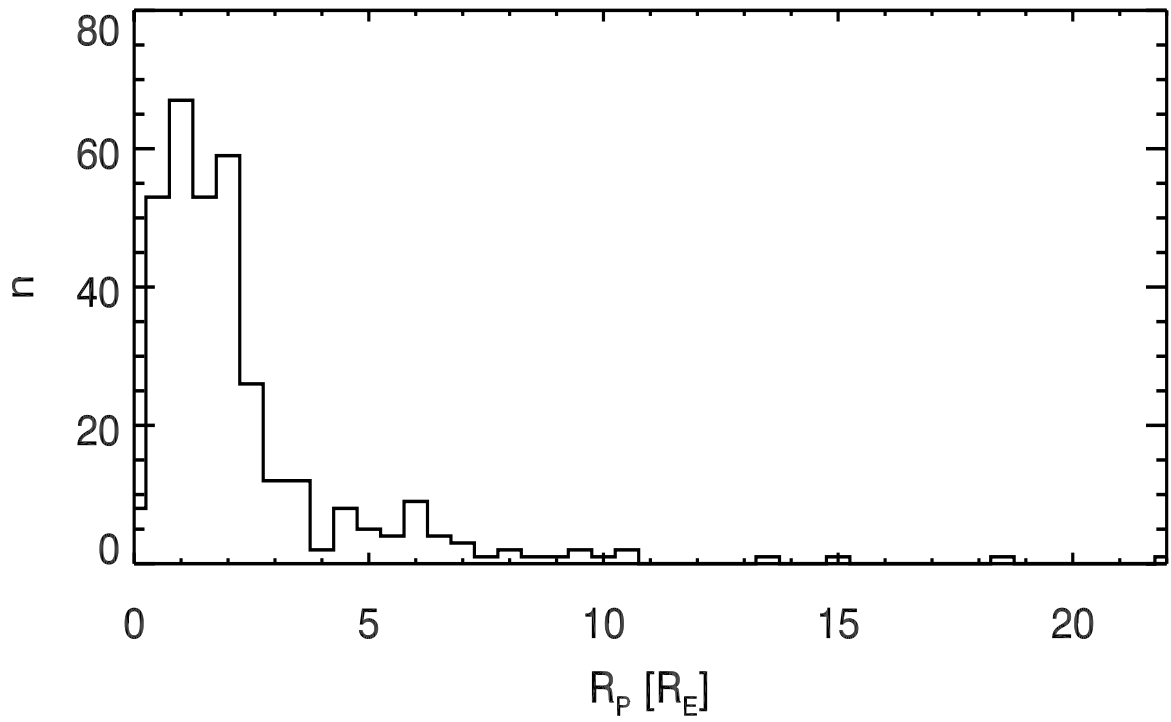} \caption{Distribution of planet radii from the iterative method for the bright multi-planet candidate systems from $Kepler$. Stellar and orbital properties can be found in Table \ref{tab:stellar_params} and Table \ref{tab:orbital_params}. 
\label{fig:Rp_hist}}
\end{center}
\end{figure}

The iterative algorithm unifies the stellar evolution modeling and the transit light curve analysis, resulting in a consistent set of stellar and orbital parameters. The results of stellar and orbital parameters for the systems in our sample are presented in Table \ref{tab:stellar_params} and Table \ref{tab:orbital_params}. Fig. \ref{fig:Kepler_Iterative_comp} compares the results of the iterative algorithm with the values from the KOI table. The iterative results generally agree well with the KOI values, as the data points are mostly located along the 1:1 line within the error bars. However, for cases when independent stellar evolution modeling and the transit light curve analysis show discrepancies, the iterative algorithm will give a different result from the KOI value because it considers measurement errors of both methods and thus gives a more consistent result. As a general trend, we found that stellar radii from KOI values are underestimated, which is in agreement with previous studies~\citep{Brown2011, Verner2011, Plavchan2013}.~\citet{Everett2013} suggested that 87\% of the KOIs' radii need an upward correction and 26\% of the KOIs' radii are underestimated by more than a factor of 1.35. In our case, we find that 86\% have radii larger than KOI values and 17\% are at least 1.35 larger than KOI values. Fig. \ref{fig:Rp_hist} shows the distribution of revised planet radii for the sample of bright $Kepler$ multiple planet candidates. The sample is dominated by smaller planet candidates. There are 233 planet candidates with radii smaller than 2.5 earth radii, 68 with radii between 2.5 and 5.0 earth radii, and 42 larger than 5.0 earth radii. 

\subsection{Search For Stellar Companions Using UKIRT Archival Images}
\label{sec:ukirt}

The UKIRT data archive in Edinburgh provides positions and magnitudes for $J$ band sources within the $Kepler$ field. Archive tools generate high resolution cut-out images on the fly and allow source cross matches around user-specified lists of coordinates. The current UKIRT dataset covers 99.5\% of the field and was observed and supplied by Phil Lucas. The images have a typical spatial resolution of 0.8-0.9 arcsec. They are therefore useful for separating blended stellar pairs and spatially resolving external galaxies.

We used UKIRT images to calculate brightness contrast curves and to detect stellar companions around planet candidate host stars. For each star in our sample, we downloaded a UKIRT image from the WFCAM Science Archive\footnote{http://surveys.roe.ac.uk/wsa/}. We calculated the median and the standard deviation of brightness (with outlier rejection) for many concentric annuli centering at the planet host star. The brightness ratio of 5-$\sigma$ above the median and the central star is defined as contrast. A contrast curve plots contrast as a function of the radii of concentric annuli, or the distance to the central star in arcsec. The contrast curve defines a brightness limit above which a visual companion can be detected. If there was a source with a brightness of at least 5$\sigma$ brighter than the median brightness, then we recorded a detection of a visual companion to the central star. 

\begin{figure}
\begin{center}
\includegraphics[angle=0, width=1.0\textwidth]{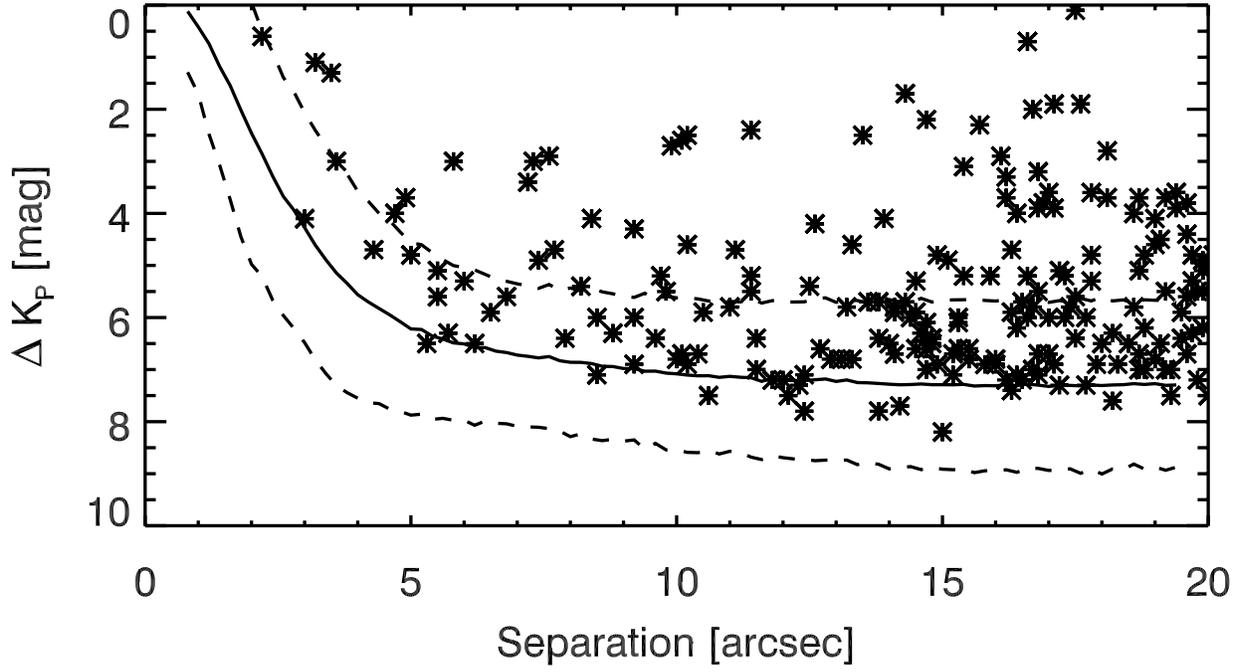} 
\caption{Averaged contrast curve for the UKIRT $J$ band images. Dashed lines are 3-$\sigma$ deviation of the contrast curve. Asterisks are detections of visual companions. A total of 177 visual companions within 20 arcsec of planet host stars are detected using UKIRT images and they are reported in Table \ref{tab:AO_params}. 
\label{fig:UKIRT_contrast}}
\end{center}
\end{figure}

Within a 20 arcsec radius centered on the central star, we detected 177 visual companions around 99 planet host stars. The other 39 stars in our sample have no visual companions down to the contrast limit. Table \ref{tab:AO_params} presents the results of visual companions detected with the UKIRT images. The average contrast curve and its 3-$\sigma$ variation are plotted in Fig \ref{fig:UKIRT_contrast}. The detections are also plotted in Fig. \ref{fig:UKIRT_contrast} as asterisks, which agree well with the average contrast curve. In this sample of 138 stars, we do not find any stars with a visual companion within 2 arcsec. We note that KOI 119 and KOI 2311 have elongated PSFs; however, the possible stellar companions failed the 5-$\sigma$ detection criteria. About 7\% (10/138) have a visual companion within 6 arcsec. The numbers are much lower than those numbers from~\citet{Adams2012} by a factor of 9. This result implies the incompleteness of visual companion searches using the UKIRT archival images, and the strength of AO observations in spatial resolution and contrast limit. However, the UKIRT images provide a contrast curve for each star and will be used to constrain parameter space in planet confidence estimation in \S \ref{sec:confidence}. There are 177, 82, 32, and 5 stars with companion star separations less than 20, 15, 10, and 5 arcsec, respectively. These numbers are roughly proportional to the sky area around the central star, implying that they are likely to be background stars rather than physically bound companions.  

\subsection{Planet Confidence}
\label{sec:confidence}

The likelihood of a planet candidate being a bona fide exoplanet can be estimated by calculating the probability of false positives and the probability of a true planet transiting event. The BLENDER technique has been used to validate $Kepler$ planet candidates~\citep[e.g., ][]{Torres2011, Fressin2011}. A similar approach was later developed for planet candidate validation~\citep{Wang2013, Barclay2013}. In order to perform planet confidence calculation, several observational constraints should be taken into consideration to exclude unlikely regions in a $\Delta K_P$-separation parameter space. $\Delta K_P$ is the differential magnitude of a possible contaminating object, and separation is its projected distance from the central star. These observational constraints include pixel centroid offset, transit depth, and UKIRT contrast curve.

The $Kepler$ light curve files contain information on both the position of the star at a given time (flux-weighted centroids, called MOM\_CENTR1 and MOM\_CENTR2) and the predicted position of the star based on the position of reference stars (called POS\_CORR1 and POS\_CORR2). By subtracting one from the other we can find the centroid position of a star. We measured the centroid position during the prospective transit, and immediately before and after transit to search for centroid shifts. If the transit occurs on a star other than the target star, the centroid position should move away from that star by an amount proportional to the transit depth~\citep{Bryson2013}. The magnitude of the shift can be used to calculate how far away this source is from the target. Even if no obvious shift is detected, we can still calculate a confusion region around the star. In our planet confidence calculations, we used the 3-$\sigma$ confusion radius as the outer limit on the separation between the target and a false positive source. 

The KOI table gives results of pixel centroid offset measurement. We used values from the flux-weighted method because this method is similar to our independent analysis. Out of 343 planet candidates in our sample, there are 240 with measured pixel centroid offset values from the KOI table, 34 of them show significant (above 3-$\sigma$) pixel centroid offsets. In comparison, for these 240 planet candidates, our independent pixel centroid offset analysis found 20 candidates with significant offsets, 14 of them are overlapped with systems characterized by significant offsets in the KOI table. The difference between our independent analysis and the measured offsets from the KOI table may be attributed to a different window size and de-trending functional form selected for analysis. For the polynomial functional form that we used for de-trending the centroid measurements, the optimal polynomial order depends on the variability of the measurement results, and the window sizes for both in- and out-of-transit measurement. However, we chose a second-order polynomial function and fixed window sizes to serve the purpose of a streamline the data reduction. These choices are empirical and may be responsible for the differences from the values from the KOI table. For the 103 planet candidates without measured pixel centroid offset values from KOI table, we found 7 with significant offsets. In the column of pixel centroid offset significance in Table \ref{tab:orbital_params}, we marked values larger than 3-$\sigma$ in bold text.  

Transit depth places a lower brightness limit to a possible contaminating object. The contrast curve of each star is calculated as described in \S \ref{sec:ukirt}. Once the observational constraints are put on the parameter space, the planet confidence will be calculated based on a galactic stellar population model~\citep{Robin2003}. The details of planet confidence calculations can be found in ~\citep{Wang2013, Barclay2013}. Table \ref{tab:orbital_params} gives the planet confidences for each of the KOIs in our sample. Two sets of planet confidences are given. The p1 column in Table \ref{tab:orbital_params} contains the planet confidences before considering the existence of other candidates in the same system. The p2 column contains the augmented planet confidences for a multi-planet system~\citep{Lissauer2012, Barclay2013}. 

\begin{figure}
\begin{center}
\includegraphics[angle=0, width=1.0\textwidth]{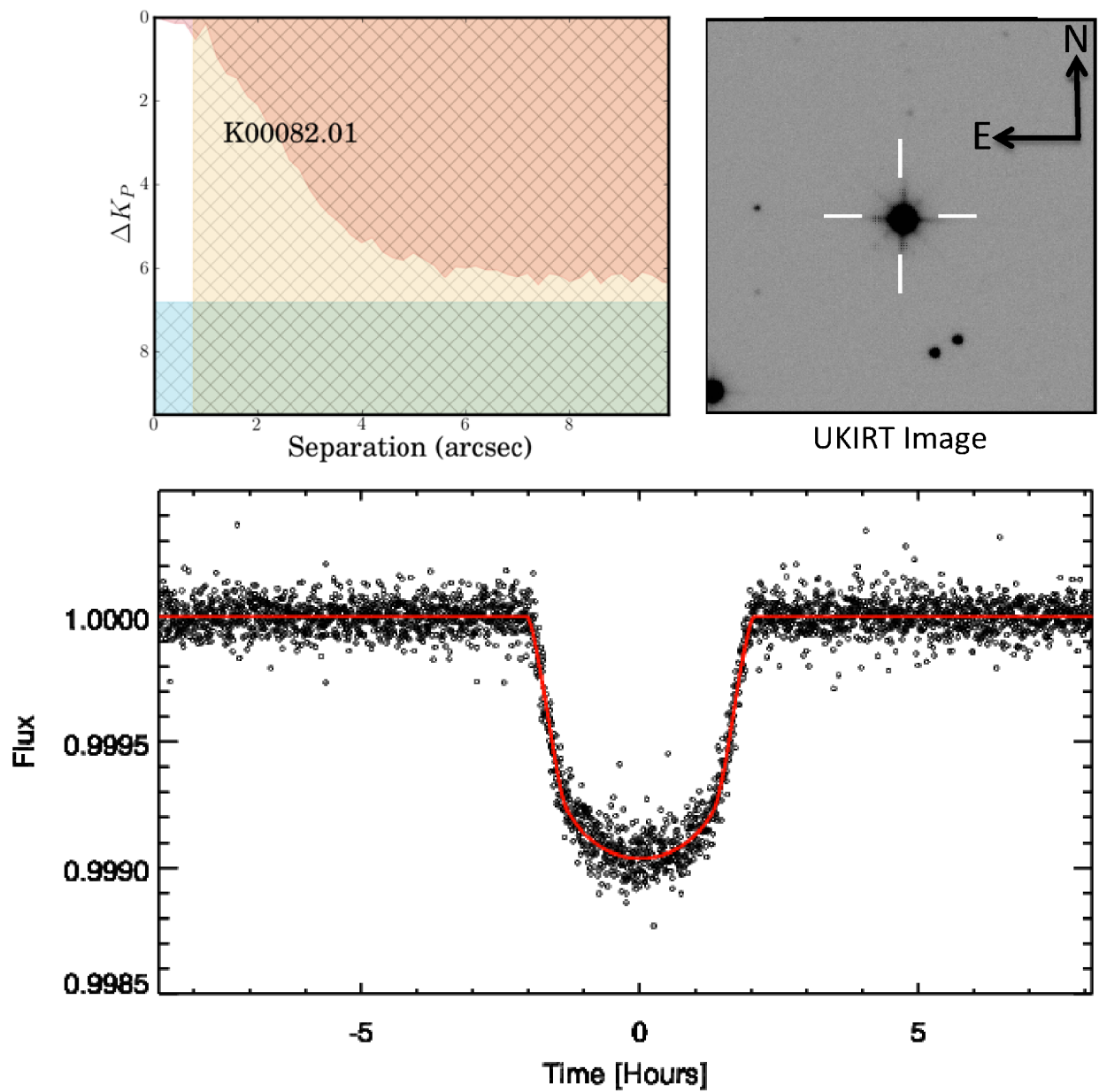} 
\caption{Top left: parameter space ($\Delta K_P-a$) for a possible contamination source around KOI 82. Yellow region is excluded by pixel centroid offset analysis, red region is excluded by contrast curve and cyan region is excluded by transit depth analysis. The remaining white region is the possible parameter space for a contamination source. Top right: the UKIRT image for KOI 82. Bottom: folded light curve for KOI 82.01(black open circles) with the best-fit transit model (red solid line). 
\label{fig:ValidatedCandidate_00082}}
\end{center}
\end{figure}

\begin{figure}
\begin{center}
\includegraphics[angle=0, width=1.0\textwidth]{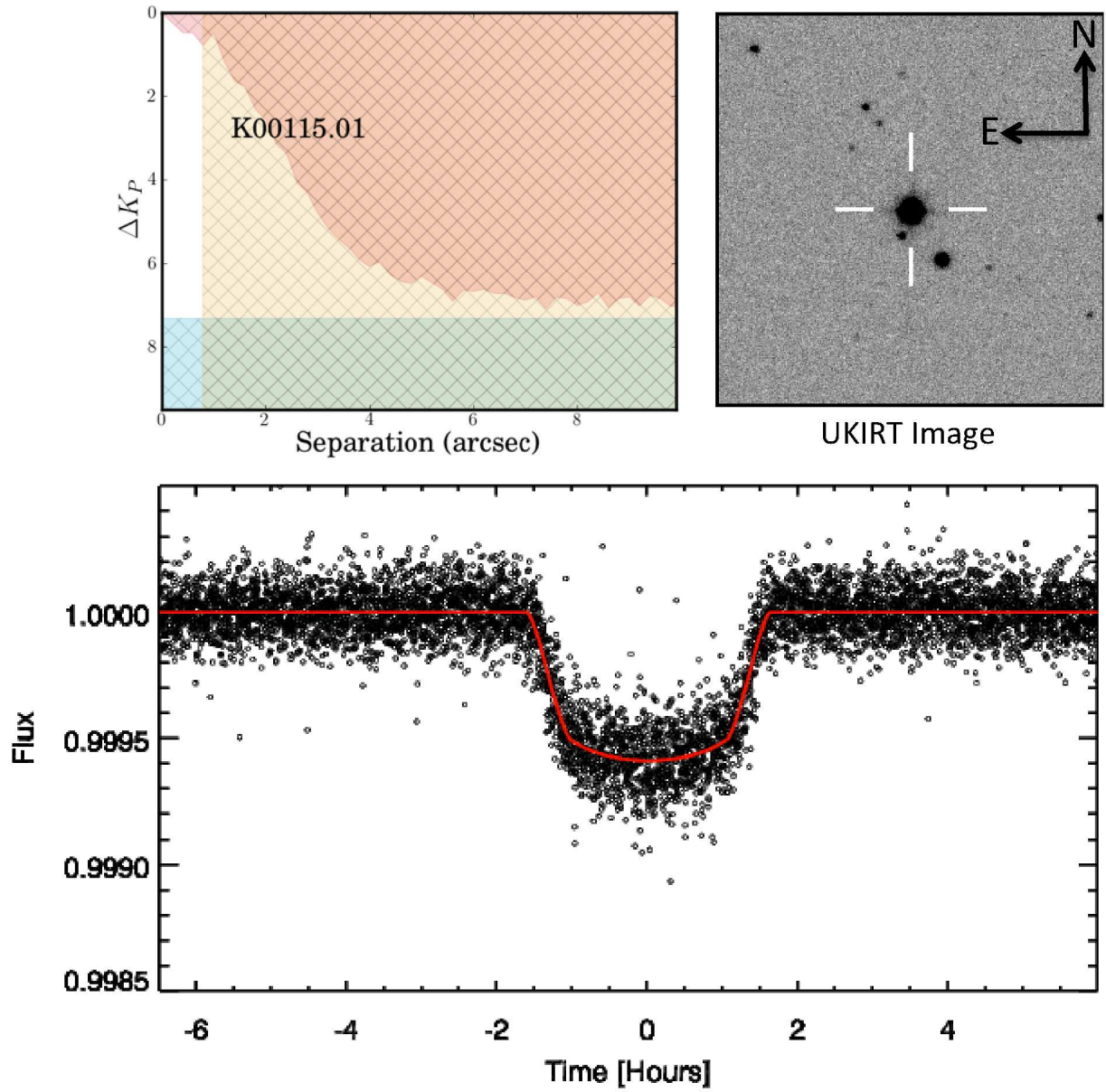} 
\caption{Same as Fig. \ref{fig:ValidatedCandidate_00082} but for KOI 115.01. 
\label{fig:ValidatedCandidate_00115}}
\end{center}
\end{figure}

\begin{figure}
\begin{center}
\includegraphics[angle=0, width=1.0\textwidth]{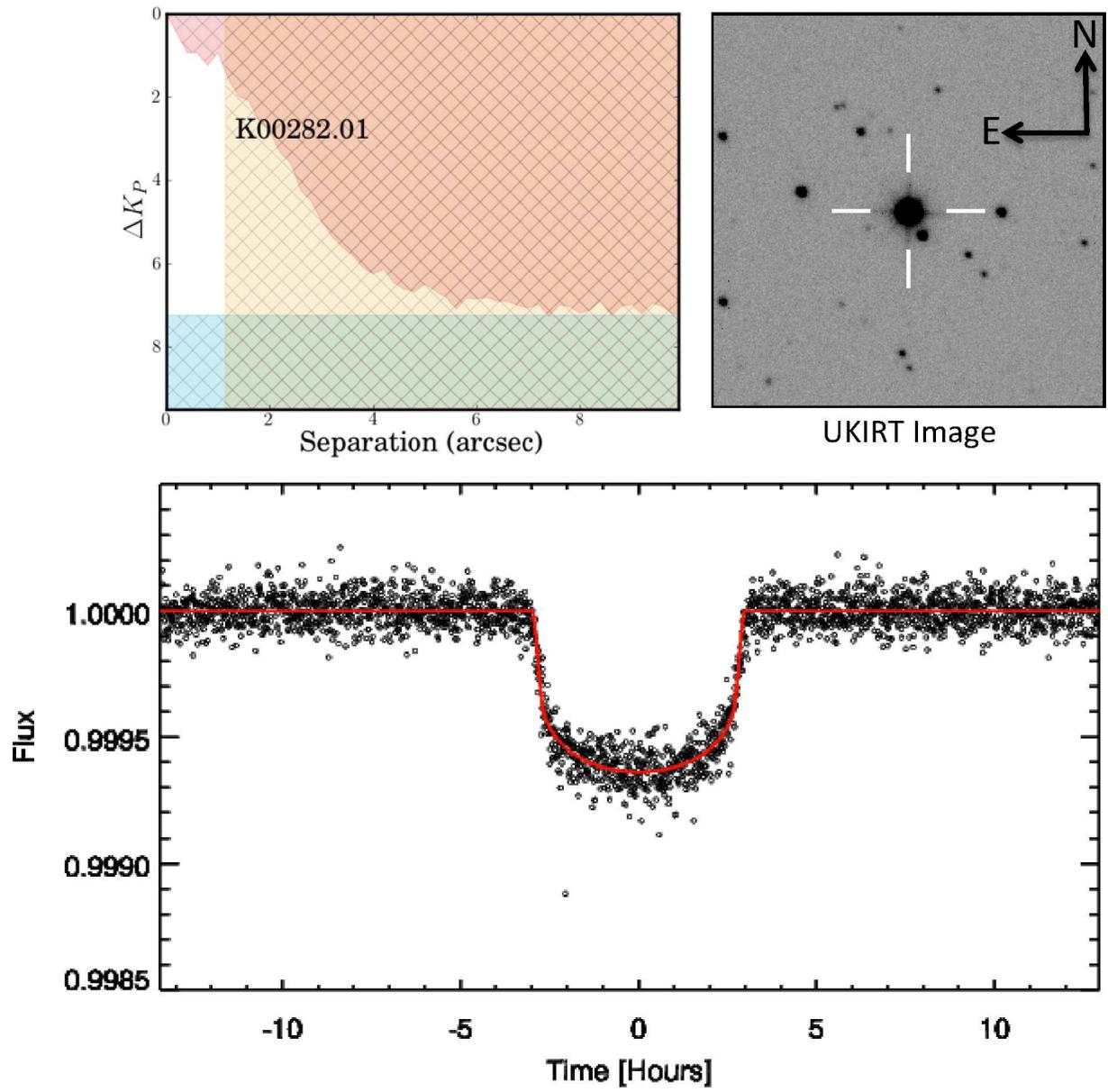} 
\caption{Same as Fig. \ref{fig:ValidatedCandidate_00082} but for KOI 282.01.
\label{fig:ValidatedCandidate_00282}}
\end{center}
\end{figure}

\begin{figure}
\begin{center}
\includegraphics[angle=0, width=1.0\textwidth]{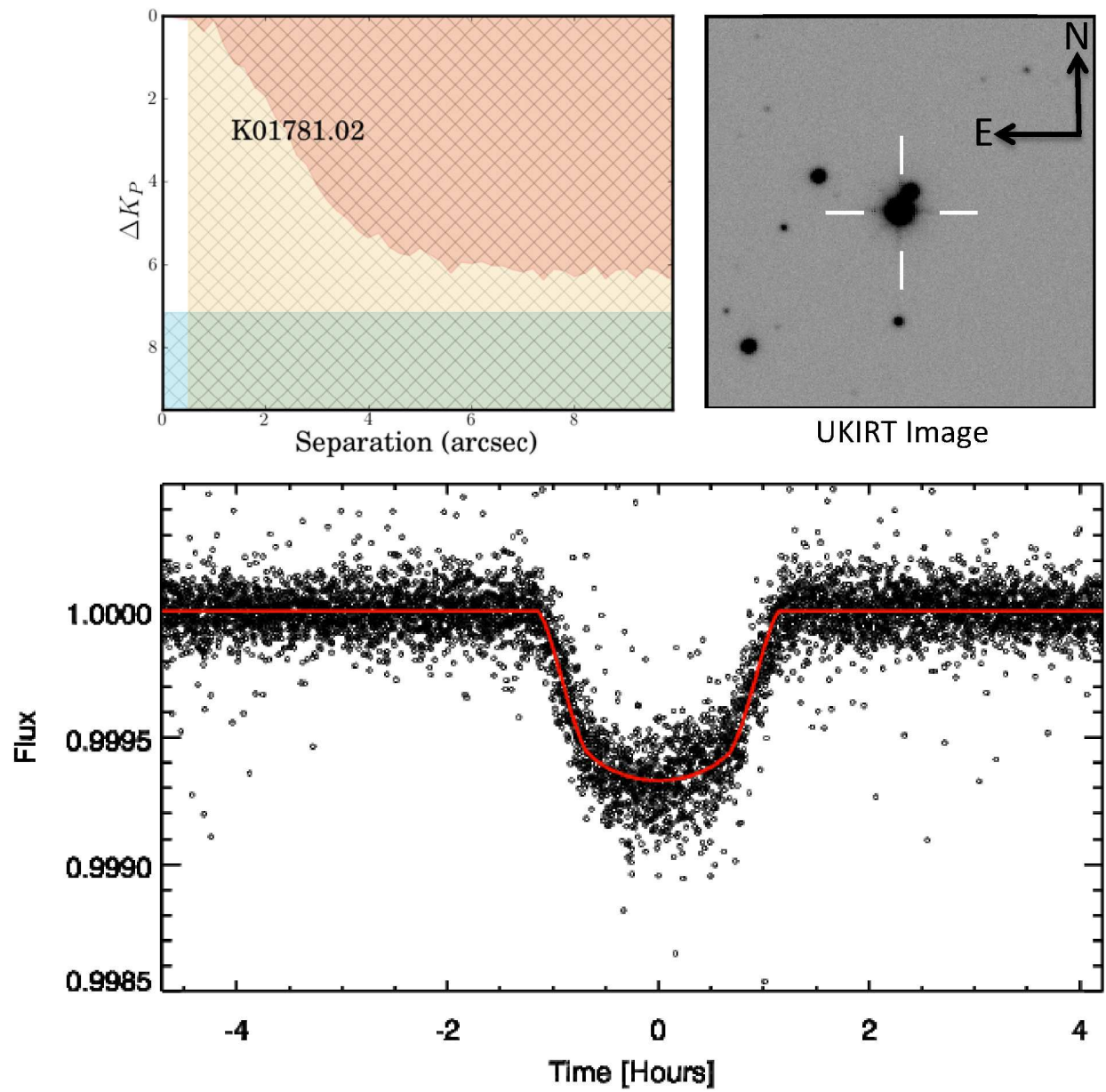} 
\caption{Same as Fig. \ref{fig:ValidatedCandidate_00082} but for KOI 1781.02.
\label{fig:ValidatedCandidate_01781}}
\end{center}
\end{figure}

There are 14 planet candidates with p2 higher than 0.997, a value high enough to promote a planet candidate to a validated planet. KOI 70.01~\citep{Fressin2012}, KOI 72.01~\citep{Batalha2011}, KOI 85.01~\citep{Chaplin2013}, KOI 94.01~\citep{Weiss2013}, KOI 244.01, KOI 244.02~\citep{Steffen2012}, KOI 245.01~\citep{Barclay2013b}, and KOI 246.01~\citep{Gilliland2013} are planet candidates that were previously validated or confirmed. For these KOIs, we note that all but KOI 70.01 have pixel centroid offsets larger than 3-$\sigma$. KOI 72, KOI 85, KOI 244, KOI 245 and KOI 246 are bright stars in the $Kepler$ field that cause pixel saturation, resulting in a imprecise pixel centroid offset measurement. KOI 94 is in a crowded field with 4 visual companions detected within 20 arcsec. We therefore caution that the flux-weighted pixel centroid offset measurement may not be precise for stars brighter than K$_P$=11, or for stars in a crowded field. There are 10 stars (7.2\%) in the sample that are brighter than 11th $Kepler$ magnitude and 9 stars (6.5\%) with more than 3 detected stellar companions within 22 arcsec. In addition to previously confirmed KOIs, we find 6 other KOIs with higher than 0.997 planet confidence: KOI 5.01, KOI 82.01, KOI 115.01, KOI 282.01, KOI 1781.01 and KOI 1781.02. However, KOI 5.01 and KOI 1781.01 have a significant pixel centroid offset, and their K$_P$s are 11.6 mag and 12.2 mag, which raises concerns about their planet nature. Therefore, our planet confidence calculation results in 4 newly validated KOIs: KOI 82.01, KOI 115.01 KOI 282.01 and KOI 1781.02. The validation plots for them are shown in Fig. \ref{fig:ValidatedCandidate_00082}, Fig.\ref{fig:ValidatedCandidate_00115}, Fig. \ref{fig:ValidatedCandidate_00282}, and Fig. \ref{fig:ValidatedCandidate_01781}. We note that KOI 115, KOI 282 and KOI 1781 all have one visual stellar companion within 8.5 arcsec. Although the possibility of the transit signal originating from the companion is excluded, flux contamination may affect the determination of the planet radius. However, given the relatively large magnitude difference between the companion and the center star ($\Delta K_P\geq 3.0$ ), the major contribution to the uncertainty of the planet radius is the stellar radius estimation rather than the flux contamination. We note that the planet confidence considers only scenarios in which a planet is physically associated with the primary. There is a chance that it orbits an undetected physical companion. This chance is likely low but is not accounted for.  

\subsection{Doppler Measurements and Incompleteness}
\label{sec:cfop}

There are 23 KOIs in our sample with more than 2 epochs of RV measurement results. They are available through the $Kepler$ Community Follow-up Observing Program (CFOP). The purpose of the website is to facilitate collaboration on follow-up observing projects of KOIs and optimize the use of available facilities. Most of the RV measurements were conducted using the Keck HIRES or SAO TRES spectrographs. Table \ref{tab:rv} summarizes the RV data for 23 KOIs. The median measurement precision is 7.3 m $\rm{s}^{-1}$ and the median observation baseline is 447.8 day. Among 23 KOIs, one non-transiting gas giant planet (Kepler-68d) was revealed by the RV technique~\citep{Gilliland2013}, and two (KOI 5 and KOI 148) exhibit a long-term RV trend. KOI 148~\citep[$Kepler$-48,][]{Steffen2013, Xie2012} shows a RV slope of $\sim$80 m $\rm{s}^{-1}\ \rm{yr}^{-1}$, but this slope is caused by a $\sim$2 Jupiter-mass planet with an orbital period of $\sim$1000 days (Howard Isaacson, private communication). KOI 5 shows a RV slope of $\sim$100 m $\rm{s}^{-1}\ \rm{yr}^{-1}$. If the mass of this companion is in the stellar regime, it must be more than 9.5 AU away from the central star in an edge-on orbit with respect to an observer. An edge-on orbit for a potential stellar companion is a probable configuration for stars with multiple transiting planets (see detailed discussions in \S \ref{sec:combination}). We therefore flag KOI 5 as a possible multiple stellar system. However, close-in (a$\leq\sim$10 AU) stellar companions are excluded. 

\begin{figure}
\begin{center}
\includegraphics[angle=0, width=1.0\textwidth]{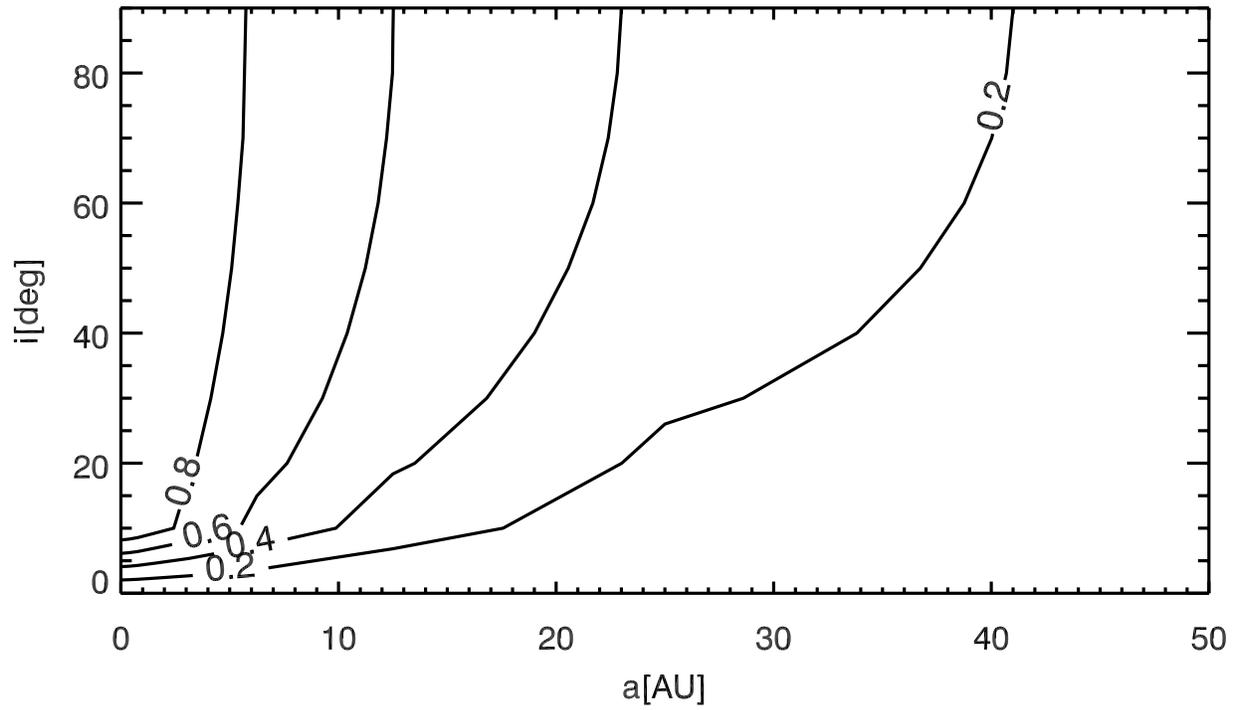} 
\caption{Averaged RV completeness contours for 23 systems in our sample with Doppler measurements. The Doppler measurements from CFOP are reported in Table \ref{tab:rv}. The Doppler technique is sensitive to companions at high inclinations and short separations. 
\label{fig:RV_completeness}}
\end{center}
\end{figure}

We studied the completeness of Doppler measurements based on the reported RV precision and observation cadence. We defined a parameter space of orbital separation and companion inclination. For each point in the parameter space, we generated a set of simulated RV signals at the reported observation epochs, but with random mass and argument of pariastron ($\omega$). The mass ratio follows a distribution of stellar companion mass ratios from ~\citet{Duquennoy1991} with a median of 0.23 and a standard deviation of 0.46. The distribution of $\omega$ is uniform between 0 and 2$\pi$. Eccentricity is assumed to be 0.4 for all simulations. This value roughly corresponds to the median eccentricity determined in studies of stellar companions around solar-type stars~\citep{Duquennoy1991, Raghavan2010}. If the the simulated RV set has a RMS scatter 3 times larger than the reported RV measurement error, then a stellar companion in a certain range of parameter space can be detected. One hundred simulations were run at each point in a grid of orbital separation and inclination space. The fraction of detectable trials is reported in Fig. \ref{fig:RV_completeness}. Completeness decreases with increasing separation and decreasing inclination. This information will be used in assessing the probability of multiple stars in calculations of the stellar multiplicity rate and the planet occurrence rate in \S \ref{sec:planet_frequency}.

\subsection{Dynamical Analysis}
\label{sec:dynamical}
A companion star can affect the stability of a planetary orbit. Given the semimajor axis of the outmost planet's orbit ($a_{\rm p}$), we have a critical (lower limit) semimajor axis of the companion \citep{HW99}, 
\begin{eqnarray}
a_{\rm c1} = a_{\rm p}/(0.464-0.380\mu-0.631e_{\rm B}+0.586\mu e_{\rm B} + 0.150e_{\rm B}^{2}-0.198\mu e_{\rm B}^{2}),
\label{ac1}
\end{eqnarray}
where $e_{B}$ is the orbital eccentricity of the stellar companion and $\mu = m_{\rm B}/(m_{\rm A}+m_{\rm B})$ is the ratio between the mass of the companion and the total mass of the stellar system.
Note that $a_{\rm c1}$ is only an empirical estimate based on the test particle simulation. Considering the mutual perturbation of planets, the real lower limit should be larger than $a_{\rm c1}$. In any case, $a_{\rm c1}$ can be treated as a lower limit.

Besides stability, a stellar companion can affect the coplanarity of a multiple planet system if the companion is on an inclined orbit with respect to the planets' orbits. This occurs when the planet's orbital node precession is dominated by stellar-planet interaction rather than the planets' mutual interaction. For a two-planet system, the node precession timescale of the outer planet due to the stellar companion $t_{\rm Bp}$ can be estimated as twice the Kozai time-scale~\citep{Kiseleva1998}, i.e., 
\begin{eqnarray}
t_{\rm Bp}\sim \frac{4}{3\pi}\frac{P_{\rm B}^{2}}{P_{\rm 2}}(1-e_{\rm B}^{2})^{3/2}\frac{m_{\rm A}+m_{\rm B}}{m_{\rm B}}
\label{tbp}
\end{eqnarray}
where $P_{\rm B}$ and $P_{\rm 2}$ are the orbital period of the stellar companion and the outer planet, respectively. The precession timescale due to planet-planet interaction can be estimated using the second order Laplace-Lagrange secular theory \citep{MD99}, i.e., 
\begin{eqnarray}
t_{\rm pp}\sim \frac{4P_{\rm 2}}{b_{3/2}^{(1)}(\alpha)} \frac{m_{\rm A}}{\alpha m_{1}+\alpha^{1/2} m_{2}},
\label{tpp}
\end{eqnarray}
where $m_{1}$ and $m_{2}$ are the masses of the two planets, $\alpha=a_{1}/a_{2}$ is the ratio of their semimajor axes, and $b_{3/2}^{(1)}(\alpha)$ is the Laplace coefficient.  
 Equating the above two timescales for precession due to stellar-planet interaction and planet-planet interaction, we obtain a critical orbital period for the companion, i.e., 
 \begin{eqnarray}
 P_{\rm B} = P_{\rm 2}\left(\frac{3\pi}{b_{3/2}^{(1)}(\alpha)}\right)^{1/2}\left(\frac{m_{\rm A}}{\alpha m_{1}+\alpha^{1/2} m_{2}}\right)^{1/2}\left(\frac{m_{\rm B}}{ m_{\rm A}+ m_{\rm B}}\right)^{1/2}\left(1-e_{\rm B}^{2}\right)^{-3/4},
 \label{pb}
 \end{eqnarray}
corresponding  to a critical semimajor axis, i.e.,
 \begin{eqnarray}
 a_{\rm c2} = a_{\rm 2}\left(\frac{3\pi}{b_{3/2}^{(1)}(\alpha)}\right)^{1/3}\left(\frac{m_{\rm B}}{\alpha m_{1}+\alpha^{1/2} m_{2}}\right)^{1/3}\left(1-e_{\rm B}^{2}\right)^{-1/2}.
 \label{ac2}
 \end{eqnarray}
For a system with more than two planets, we calculate equation \ref{ac2} for each pair of planets and adopt the minimum value as the $a_{\rm c2}$ of the system. Below this critical semi-major axis, planetary coplanarity would be significantly affected by the companion's perturbation. 
Generally, $a_{\rm c2}>a_{\rm c1}$ for a typical planet system, and thus coplanarity puts a stronger constraint on the companion's orbit. 

\begin{figure}
\begin{center}
\includegraphics[angle=0, width=1.0\textwidth]{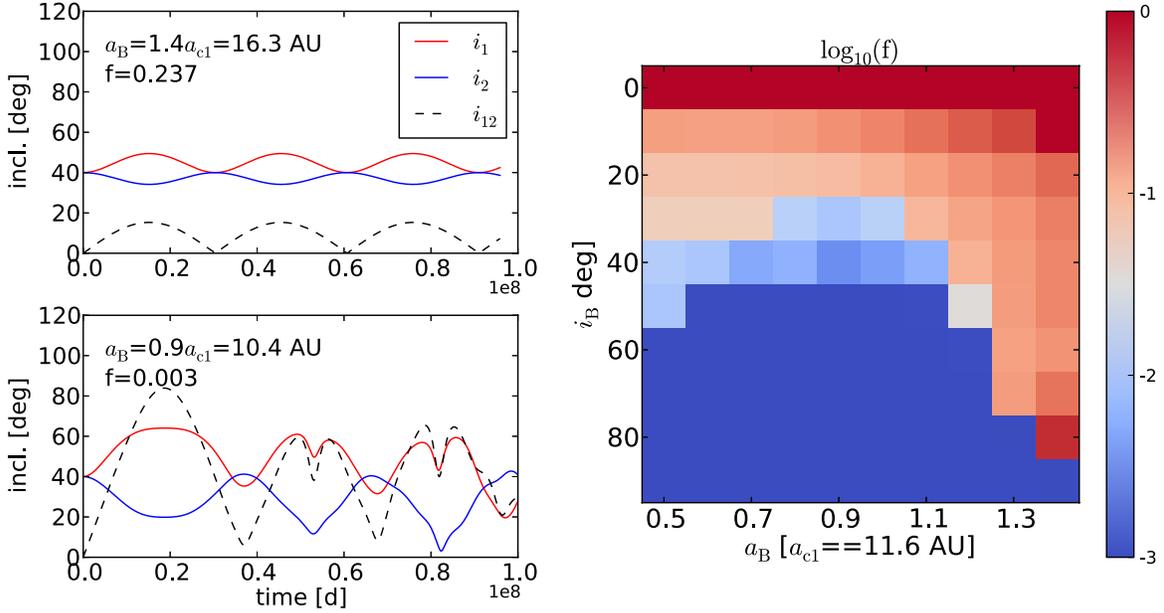} 
\caption{An example of dynamical analysis for KOI 275 system, which
has two planet candidates KOI-275.01 and KOI 275.02. The left two
panels show the evolutions of their orbital inclinations relative to
the orbit plane of the proposed stellar companion ($i_1$ and $i_2$) and
their mutual orbit inclination ($i_{12}$) in the cases of $a_B$ = 16.3 AU
(top) and $a_B$ = 10.4 AU (bottom). In both cases, the initial orbital
inclination of the binary relative to the planet candidates is set to
$i_B$ = 40$^\circ$. These two cases correspond to two grids in the right
panel, which maps the time fraction (f) that both planet candidates
stay in a coplanar configuration with $i_{12}$ $<$ 5$^\circ$ (f = 0.237 and f = 0.003 for the two cases shown in the left panels, respectively). Since
orbital plane of a transiting planet is almost perpendicular to the
sky plane, and thus here $i_B$ = 90$^\circ$ - $i$, where $i$ is the orbital
inclination with respect to the sky (used in Fig. \ref{fig:RV_completeness} and \ref{fig:RV_Dynamical_completeness}) 
\label{fig:DA_illustration}}
\end{center}
\end{figure}

Our analysis gives a qualitative estimate of how planetary stability and coplanarity would be affected by a stellar companion. To quantitatively measure such effects, we perform the following an N-body simulation. For each multiple KOI system, we add a test stellar companion and use the N-body simulation package MERCURY \citep{Cha97} to integrate the orbits of the system up to a timescale of $10t_{\rm Bp}$. In each integration, we monitor the relative orbital inclinations of all the planets and calculate the fraction of time during which their relative inclinations are all less than $5^{\circ}$. This time fraction is then treated as the probability\footnote{We set $\rm P_{DA}=1$ if there was any instability event, e.g., ejection or collision.}($\rm P_{DA}$) at which the proposed stellar companion can not be ruled out. Here, the proposed stellar companion is drawn from a range of semimajor axes of $0.5a_{\rm c2}<a_{\rm B}<2a_{\rm c2}$ and a range of inclinations of $ 0 <i_{\rm B}< 90^{\circ}$. The companion's mass and orbital eccentricity are set to: $m_{\rm B}= 0.1 M_{\odot}$ and $e_{\rm B}=0.1$, to get a conservative estimate of $\rm P_{DA}$. The relative inclination cut-off at $5^{\circ}$ is chosen because $Kepler$ planets are believed to be highly coplanar with a relative inclination within $1^{\circ}-2^{\circ}$~\citep{Lissauer2011,Fang2012,Tremaine2012,Fabrycky2012,Figueira2012, Johansen2012}. In all simulations, the planet mass is set to a nominal value, $m_{\rm p}=(R_{\rm p}/R_{\oplus})^{2.06} M_{\oplus}$ \citep{Lissauer2011}, and the planets are started with circular and coplanar orbits. As an example, Fig. \ref{fig:DA_illustration} plots the result for KOI 275.

\section{Planet Occurrence Rate and Stellar Multiplicity}
\label{sec:planet_frequency}

Planet host stars with detected physical companions are categorized as multiple stars. Those with non-detections are assigned a probability of being a multiple star. The probability is calculated based on the overall search completeness for the star and the stellar multiplicity rate for solar-type stars. For example, if the overall completeness for a companion detection is 80\% and the stellar multiplicity rate is 46\%~\citep{Raghavan2010}, then the probability of the star without a detected companion is (100\%-80\%)$\times$46\%=0.092. Following this procedure, we calculate the number of multiple stars $N_M$ and the number of single stars $N_S$. $N_M$ and $N_S$ are the sums of probabilities, so they will not necessarily be integers:
\begin{equation} 
\label{eq:nmns}
N_M=\sum\limits_{i=1}^{n} p_M(i),\ N_S=\sum\limits_{i=1}^{n} [1-p_M(i)],
\end{equation}
where $p_M(i)$ is the probability of the $i_{\rm{th}}$ star being a multiple star.

The ratio of planet occurrence rate for multiple stars ($f_M$) and single stars ($f_S$) can be calculated with the ratio of $N_{PM}/N_{PS}$ and the stellar multiplicity rate for the field stars:
\begin{equation} 
\label{eq:fsfm}
\frac{f_S}{f_M}=\left({\frac{N_{PS}}{0.54}}\right)/\left({\frac{N_{PM}}{0.46}}\right),
\end{equation}
where $N_{PM}$ and $N_{PS}$ are the numbers of planets around multiple stars and single stars in the sample:
\begin{equation} 
\label{eq:npmnps}
N_{PM}=\sum\limits_{i=1}^n\sum\limits_{j=1}^m p_M(i)\cdot p_{PL}(i,j),\ N_{PS}=\sum\limits_{i=1}^n\sum\limits_{j=1}^m [1-p_M(i)]\cdot p_{PL}(i,j),
\end{equation}
where $p_{PL}(i,j)$ is the planet confidence of the $j_{\rm{th}}$ planet candidate in the $i_{\rm{th}}$ star with a total of $m$ planet candidates in the system. The planet confidence is a measure of the relative probability of a bona-fide planet to false positives  after a series of false positive tests (see \S \ref{sec:confidence} for details). The overall planet occurrence rate (in the unit of number of planet per system) can be calculated by the following equation:
\begin{equation} 
\label{eq:fre_all}
f={0.54\times f_S+0.46\times f_M},
\end{equation}
where $f$ is overall planet occurrence rate. With both the stellar multiplicity rate and $f_S/f_M$ in hand, the planet occurrence rate for multiple stars and single stars can be calculated:
\begin{equation} 
\label{eq:fre}
f_M=\frac{f\cdot (\frac{0.54}{0.46}+1)}{\frac{0.54}{0.46}\cdot\frac{f_S}{f_M}+1},\ f_S=f_M\cdot \frac{f_S}{f_M}.
\end{equation}
The planet occurrence rate for single stars and multiple stars can therefore be distinguished and compared. With Equation \ref{eq:fre}, the comparison of $f_S$ and $f_M$ will lead to the understanding of the influence of stellar multiplicity on planet formation. 

\subsection{Combining Doppler and Dynamical Results}
\label{sec:combination}

The Doppler technique and the dynamical stability analysis are two complementary methods, with the former sensitive to edge-on companions and the latter sensitive to face-on companions. The difference between these two methods is that the Doppler technique sets constraints from observation while the constraints from the dynamical analysis come from numerical simulations.  

At each point in the $a$-$i$ parameter space for a possible stellar companion in a planetary system, two constraining numbers are given. One is from the Doppler analysis and the other one is from the dynamical analysis. For example, the completeness of the Doppler technique at [$a$,$i$]=[10 AU, 50$^\circ$] is 60\%, and the dynamical analysis predicts that 10\% of simulations do not meet coplanarity condition. The probability of an undetected companion is therefore $40\%\times90\%\times p(a,i)$, where 40\% is the incompleteness of Doppler observations, 90\% is the probability of coplanar configuration from the dynamical analysis, and $p(a,i)$ is the probability of a stellar companion at a given $a$ and $i$. We note that the results from Doppler observation and the dynamical analysis may be correlated. For example, the 10\% rejected orbits in the previous example may be partially detected by Doppler observations. The correlation will result in an underestimation of the probability of an undetected companion. However, the correlation should be small because they are sensitive to different phase spaces and the results are thus nearly orthogonal. Most face-on companions missed by Doppler observations will be excluded by the dynamical analysis, and vice versa. Highly eccentric orbits that are occasionally missed by Doppler observations due to limited phase coverage can also be excluded by the dynamical analysis.

We treat the probability $p(a,i)$ as a product of $p(a)$ and $p(i)$. We used the Gaussian distribution of stellar companion periods of solar type stars~\citep{Raghavan2010}, and converted periods to separations to describe $p(a)$. For $p(i)$, the distribution of inclinations for stellar companions, we adopted the result from ~\citet{Hale1994}. All systems in our sample are multi-planet systems, and there is evidence of alignment of stellar spin and planet orbital planes for multi-planet systems~\citep{Hirano2012, Albrecht2013}.~\citet{Hale1994} suggested that the spin-orbit alignment is likely for binaries with separations less than $\sim$15 AU and the spin-orbit angle becomes random when separations exceed $\sim$30 AU. Therefore, if there is a stellar companion within 15 AU around a star in our sample, its orbital plane should align approximately with the orbital plane of the planets. This alignment probability becomes smaller and randomized as the separation grows larger than 30 AU. We used different functions of $p(i)$ for different separations. For separations less than 15 AU, we adopted a Gaussian form with a median of 80$^\circ$ and a standard deviation of 5$^\circ$; for $15 \rm{AU} \leq a \leq 30 \rm{AU}$, we used a combination of the Gaussian form and a uniform distribution between 0$^\circ$ to 90$^\circ$ with equal probability; for separations larger than 30 AU, we used a uniform distribution between 0$^\circ$ and 90$^\circ$ to describe $p(i)$. These functions at different separations agree well with the observations in~\citet{Hale1994}. The probability of a star being in a multiple stellar system is then calculated by integrating over the $a$-$i$ phase space.

\subsection{Planet Host Stars Multiplicity Rate}
\label{sec:multip}

\begin{figure}
\begin{center}
\includegraphics[angle=0, width=1.0\textwidth]{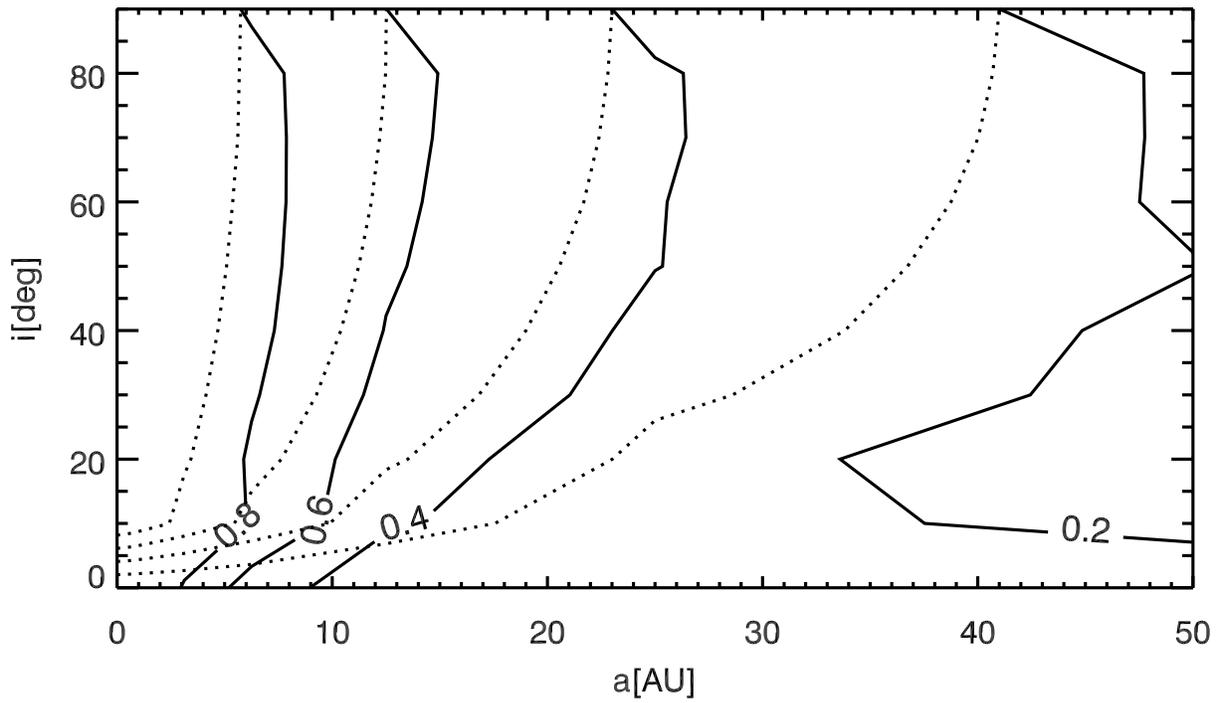} 
\caption{Averaged completeness contours for Doppler measurements and the dynamical stability analysis. Contours for RV-only completeness from Fig. \ref{fig:RV_completeness} are over plotted as dotted lines. The completeness is improved by the sensitivity to stellar companions with low $i$ values by the dynamical stability analysis. 
\label{fig:RV_Dynamical_completeness}}
\end{center}
\end{figure}

Fig. \ref{fig:RV_Dynamical_completeness} shows the average sensitivity contours after combining the Doppler observations and the dynamical analysis. When compared to Fig. \ref{fig:RV_completeness}, the Doppler observation incompleteness is complemented by the dynamical analysis at low inclinations. There are a total of 23 systems with Doppler observations among 138 systems in our sample.

Fig. \ref{fig:Multi_PlanetHost_Field} shows the comparison of the stellar multiplicity rate for field stars and planet host stars as a function of separation. We performed calculations for two samples. The first sample contained stars with both Doppler measurements and dynamical analysis (RV sample, N=23), and the second sample included all stars (N=138). The dashed line is the field star multiplicity rate~\citep{Raghavan2010}. The blue region is the calculated stellar multiplicity rate, i.e., $N_M/N$, for the RV sample, where the vertical height of the region at a given separation indicates the 1-$\sigma$ error bar. The error bar of $N_M$ is estimated based on Poisson statistics. The square root of the closest integer to $N_M$ is used to estimate the error of $N_M$ unless the closest integer is 0, in which case, we used 1 for the error of $N_M$. At $a$ = 20.8 AU, the blue hatched area crosses over the dashed line, suggesting that the stellar multiplicity rates for field stars and for the RV sample become indistinguishable beyond $\sim$20 AU. The lower stellar multiplicity rate for planet host stars suggests that the presence of a close-in stellar companion may have a suppressive effect on planet formation and evolution. From the completeness contours in Fig. \ref{fig:RV_Dynamical_completeness}, the completeness is $\sim$50\% around 20 AU. At this completeness level, the majority of stellar companions within 20 AU would be detected or excluded due to orbit stability concern, leading to a conclusion of suppressive influence of a close-in stellar companion within 20 AU. 

The red hatched area is the stellar multiplicity rate for the entire sample of multi-planet host stars. The error bar is smaller due to a larger sample. The crossover with the dashed line takes place at 85 AU. However, the significant incompleteness beyond 20 AU (see Fig. \ref{fig:RV_Dynamical_completeness}) prevents us from determining whether the crossover is due to an indistinguishable stellar multiplicity rate between planet host stars and field stars, or merely an incomplete survey. If it is the former case, then the large sample helps to reduce the statistical error and push the effective separation of a stellar companion from 20 AU to 85 AU. If it is the latter case, a longer RV measurement baseline or AO imaging would detect more stellar companions, pushing the red hatched area upward and the effective separation closer than 85 AU. Nonetheless, the suppressive influence of a close-in stellar companion is shown. It is the effective separation that is currently lacking a satisfactory constraint due to statistical error and survey incompleteness.

\subsection{Planet Occurrence Rate For Single and Multiple Stars}
\label{sec:fre}

\S \ref{sec:planet_frequency} provides a method of calculating the planet occurrence rate for single and multiple stars respectively. The result depends on $f$, the overall planet occurrence rate. Since $f$ is still an issue of debate~\citep{Cumming2008, Howard2010, Mayor2011, Catanzarite2011, Youdin2011, Traub2012, Howard2012, Mann2012, Gaidos2013, Swift2013, Kopparapu2013, Bonfils2013, Fressin2013, Dressing2013, Parker2013, Petigura2013, Petigura2013b}, we will leave it in the equation as a variable to be determined. Planet occurrence rate calculations need to cover a wide range of separations to avoid omitting any possible companions. The distribution of separations for stellar companions approaches zero when $a$ gets to more than $10^5$ AU, therefore we truncate the $a$ range at $10^5$ AU. We get the ratio of $f_S$ and $f_M$ very close to 1, indicating that a planet is equally probable to appear around a single star or around a multiple star system. This result is not surprising due to the incompleteness of  Doppler observations and the dynamical analysis. There is a large unexplored phase space as the calculation covers a wider separation range. The prior information, e.g., the stellar multiplicity rate for the field star, outweighs the observations and analysis, naturally leading to a conclusion purely based on priors of the calculation. 

We could potentially reduce the separation range and therefore reduce the impact of priors. However, this would also reduce the value of this analysis. Followup observations that address the unexplored phase space are the only way to reach a sensible planet occurrence rate estimation for both single and multiple stars. It requires a longer time baseline of Doppler observations for detection of close-in companions and imaging techniques with higher spatial resolutions (e.g., AO, speckle imaging and HST snapshot program\footnote{http://www.stsci.edu/cgi-bin/get-proposal-info?id=12893\&submit=Go\&observatory=HST}) to probe for companions at wide orbits with deeper contrast. 

\section{Summary and Discussion}
\label{sec:Summary}

\subsection{Summary}
\label{sec:sum}

We selected a total of 138 multi-planet candidate systems from the $Kepler$ mission and studied the influence of a nearby stellar companion on planet formation. For these systems, we used archival data from $Kepler$ to characterize their stellar and orbital properties. We developed an iterative algorithm that combines the $Kepler$ photometric data with a stellar evolution interpolator. This algorithm uses $a/R_S$ as a bridge to obtain a self-consistent stellar and orbital solution. We used this algorithm to refine stellar and orbital solutions for the 138 systems in our sample. The results are reported in Table \ref{tab:stellar_params} and Table \ref{tab:orbital_params}. The majority of the planet candidates are in compact systems: 90\%(310/343) of them have semi-major axes less than 0.28 AU. We used the archival data from UKIRT to study the stellar environment of these planet host stars, and found 177 visual companions within 20 arcsec, but we argued that these are likely to be background stars according to a statistical test. 

We calculated the planet confidence for these multi-planet candidate systems (Table \ref{tab:orbital_params}). The calculation involves pixel centroid offset analysis, transit depth analysis, and contrast curve calculations. A total of 14 planet candidates have a planet confidence higher than 0.997, indicating a high likelihood of being a bona fide planet. Eight of them are previously confirmed or validated planets. The other 6 planet candidates are validated for the first time, but two of them have a significant pixel centroid offset. Thus, KOI 82.01, KOI 115.01, KOI 282.01 and KOI 1781.02 become 4 newly validated planets from $Kepler$ data. Future AO observations will further reduce the parameter space for a possible contaminating stellar companions, and increase the planet confidences for other unvalidated planet candidates. 

The majority of previous work on stellar multiplicity made use of imaging techniques, such as AO and Lucky imaging, and searched for stellar companions at wide separations. In comparison, we used a unique combination of the Doppler technique and dynamical stability analysis to evaluate stellar companions out to $\sim$100 AU. This approach demonstrates the influence of a close-in stellar companion on planet formation. We presented a statistical method for calculating the stellar multiplicity rate and the planet occurrence rate for both single stars and multiple stars. We found that the stellar multiplicity within 20.8 AU for planet host stars is significantly lower than that of field stars, indicating a strong suppressive effect of a stellar companion on planet occurrence. The influence of a stellar companion at larger separations is uncertain because of statistical error and survey sensitivity. Doppler observations for the 23 KOIs in our sample are not optimal for a search of stellar companions at larger separations. Future followup Doppler observations with a longer time base line will help set further constraints on stellar companions. 

\subsection{Discussion}
\label{sec:dis}

In our dynamical analysis, we used several conservative assumptions. First, we assumed a moderate eccentricity of 0.1 in the analysis. If this is replaced by 0.5, roughly the median of eccentricity for binaries~\citep{Raghavan2010}, a larger fraction of orbital configurations will become non-coplanar for the same $a$ and $i$. The dynamical analysis thus becomes even more effective with a higher eccentricity assumption. Second, we assumed a mutual inclination limit to be 5$^\circ$, which is a weaker constraint compared to the highly coplanar orbits found by $Kepler$ with a mutual inclination of less than 2$^\circ$~\citep{Lissauer2011,Fang2012,Tremaine2012,Fabrycky2012,Figueira2012, Johansen2012}. A decreasing limit of mutual inclination will result in more rejections in dynamical analysis and thus increase the constraint on undetected stellar companions. However, we must note some limitations of the dynamical analysis. For example, we do see several cases in which a possible planet candidate is found in a region predicted to be unstable by the dynamical analysis, e.g., KOI 191 and 204~\citep{Lissauer2011}, and $\nu$ Octantis~\citep{Gozdziewski2013}.

\begin{figure}
\begin{center}
\includegraphics[angle=0, width=1.0\textwidth]{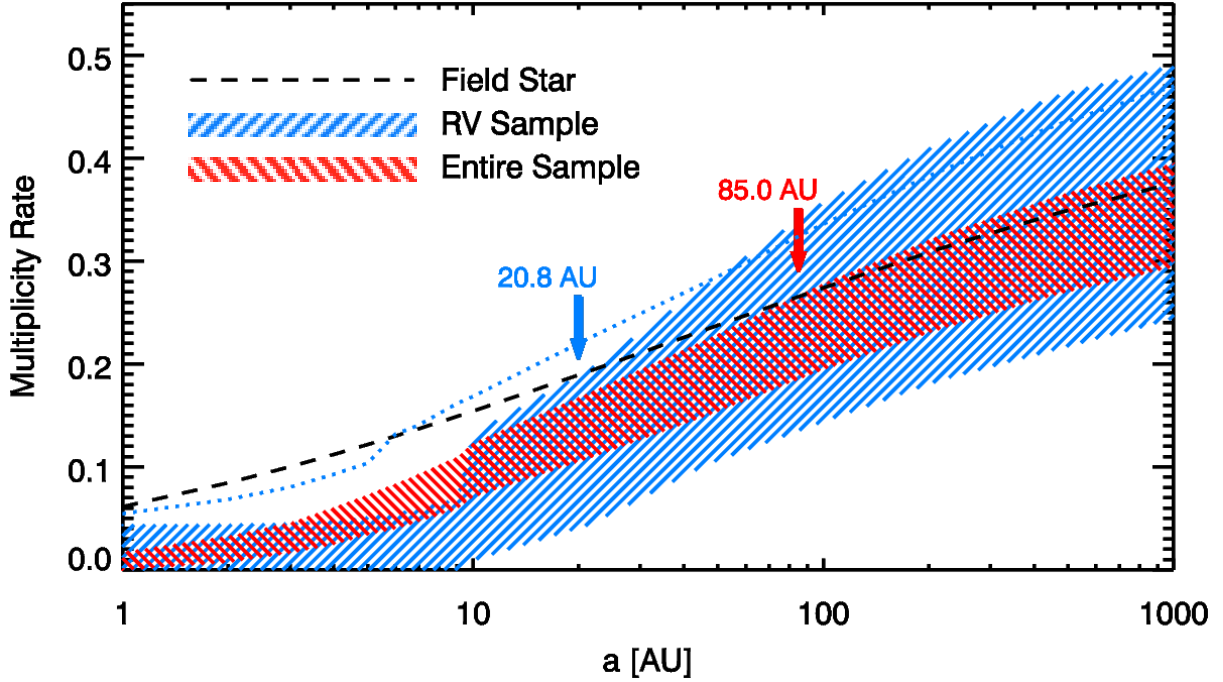} 
\caption{The comparison of field star multiplicity rate and the stellar multiplicity rate for planet host stars. Dashed line represents the stellar multiplicity rate for field stars. Blue hatched area represents 1-$\sigma$ uncertainty region of stellar multiplicity rate for 23 KOIs with both RV measurements and dynamical analyses. There is a significant deficiency of stellar companions within 20.8 AU, indicating that planet formation and evolution may be suppressed by a close-in stellar companion. If only dynamical analysis is used for the RV sample, the 1-$\sigma$ upper envelope is shown by the blue dotted line. In this case the difference between the field star multiplicity rate and that for the planet host stars is less distinct due to less constraint from observation and larger statistical errors, which emphasizes the important role of followup RV observations. Red hatched area represents 1-$\sigma$ uncertainty region of the stellar multiplicity rate for 138 KOIs with dynamical analyses including 23 KOIs with RV measurements. The red hatched area suggests a wider effective separation below which planet formation and evolution may be significantly affected, but incompleteness beyond 20 AU prevents us from concluding whether the wider effective separation is due to stellar companion perturbation or an incomplete search and exclusion for stellar companion. 
\label{fig:Multi_PlanetHost_Field}}
\end{center}
\end{figure}

The stellar multiplicity rate for field stars in the solar neighborhood has been calculated by~\citet{Raghavan2010} for solar-type stars within 25 pc of the Sun. The majority of the stars in our sample are within 550 pc of the Sun based on the magnitude cut at 13.5 $Kepler$ mag and the assumption that most of them are main sequence stars. The caveat of extrapolating the measurement to a larger volume must be noted. If the stellar multiplicity rate of the $Kepler$ field is different from the solar neighborhood, the ratio (0.54/0.46) in \S \ref{sec:planet_frequency} should be replaced with a different (but unknown) value. In addition, we compared the stellar multiplicity rate for field stars and planet host stars, but we do not know the fraction of field stars hosting planets. If all field stars have planets, then the comparison provides information about the difference of stellar companions in terms of separation distribution and the stellar multiplicity rate, because we would have compared two groups of planet hosting stars in two environments. One is the solar neighborhood, the other is the $Kepler$ field. However, if (1) not all field stars have a planet; and (2) the statistics of multiple stars (multiplicity, separation distribution, etc.) is similar for the nearby solar-type stars and the stars in our sample, then the difference in Fig. \ref{fig:Multi_PlanetHost_Field} indeed suggests the impact of a close-in stellar companion on planet occurrence. In this case, the field stars are a sample contaminated by planet host stars. If a difference is seen when compared to a sample of planet host stars, then the difference would have been more distinct when comparing a planet host sample and a non-planet host sample. The latter is difficult to obtain because of current detection precision limitation. However, a planet mass or radius limit can be set in investigations of a certain type of planet, e.g., comparing the stellar multiplicity rate for the giant planet host stars and stars without a gas giant. For future studies, the combination of the Doppler technique, AO observation and dynamical stability analysis will contribute an almost complete survey for stellar companions around planet host stars,  and will ultimately reveal how stellar multiplicity influences on planet formation.


\bibliography{mybib_JW_DF_PH5}

\clearpage


\begin{deluxetable}{lccc}
\tablewidth{0pt}
\tabletypesize{\tiny}
\setlength{\tabcolsep}{0.02in}
\tablecaption{List of references on stellar multiplicity study of planet host stars.\label{tab:ReferencesList}}
\tablehead{
\colhead{\textbf{Reference}} &
\colhead{\textbf{Multiplicity}} &
\colhead{\textbf{Sample Size}} &
\colhead{\textbf{Technique}} \\

}

\startdata

~\citet{Luhman2002} & 16\% & 25 & AO \\
~\citet{Patience2002} & 27\% & 11 & AO \\
~\citet{Eggenberger2007a, Eggenberger2011} & 13.2\%$\pm$5.1\% less than that of control group & 130 & AO  \\
~\citet{Daemgen2009} & 21\% & 14 & Lucky Imaging \\
~\citet{Mugrauer2009} & 17\%  & 250 & wide field imaging \\
~\citet{Raghavan2010} & 30\% & 36 & interferometry, CPP and RV \\
~\citet{Roell2012} & $>$12\% & 477 & literature review and catalog search\\
~\citet{Adams2012, Adams2013} & 60\%, 20\% and 7\% within 6, 2 and 0.5 arcsec & 90 & AO \\
~\citet{Ginski2012} & 1.5\% & 71 & Lucky Imaging \\
~\citet{LilloBox2012} & 12\%-18\% & 98 & Lucky Imaging \\
~\citet{Bergfors2013} & $>$38\% & 31 & Lucky Imaging \\

\enddata
\end{deluxetable}

\newpage

\clearpage

\begin{landscape}
\begin{deluxetable}{lcccccccccccccc}
\tablewidth{0pt}
\tabletypesize{\tiny}
\setlength{\tabcolsep}{0.02in}
\tablecaption{Stellar Parameters\label{tab:stellar_params}}
\tablehead{
\colhead{\textbf{ }} &
\multicolumn{7}{c}{\textbf{KOI}} &
\multicolumn{7}{c}{\textbf{Derived Values}} \\

\colhead{\textbf{KOI$^{\dagger}$}} &
\colhead{\textbf{KIC$^{\dagger}$}} &
\colhead{\textbf{$\alpha$$^{\dagger}$}} &
\colhead{\textbf{$\delta$$^{\dagger}$}} &
\colhead{\textbf{K$_P$$^{\dagger}$}} &
\colhead{\textbf{$T_{\rm eff}$$^{\dagger}$}} &
\colhead{\textbf{$\log g$$^{\dagger}$}} &
\colhead{\textbf{[Fe/H]$^{\dagger}$}} &
\colhead{\textbf{$T_{\rm eff}$}} &
\colhead{\textbf{$\log g$}} &
\colhead{\textbf{[Fe/H]}} &
\colhead{\textbf{M$_{\ast}$}} &
\colhead{\textbf{R$_{\ast}$}} &
\colhead{\textbf{L$_{\ast}$}} &
\colhead{\textbf{$\rho_{\ast}$}}  \\

\colhead{\textbf{}} &
\colhead{\textbf{}} &
\colhead{\textbf{(deg)}} &
\colhead{\textbf{(deg)}} &
\colhead{\textbf{(mag)}} &
\colhead{\textbf{(K)}} &
\colhead{\textbf{(cgs)}} &
\colhead{\textbf{(dex)}} &
\colhead{\textbf{(K)}} &
\colhead{\textbf{(cgs)}} &
\colhead{\textbf{(dex)}} &
\colhead{\textbf{(M$_{\odot}$)}} &
\colhead{\textbf{(R$_{\odot}$)}} &
\colhead{\textbf{(L$_{\odot}$)}} &
\colhead{\textbf{(g~cm$^{-3}$)}} 

}

\startdata

K00005	  & 8554498    & 289.739716      & 44.647419       & 11.66 & 5861       & 4.19       & 0.12       &$ 5865  ^{+ 60    }_{ -58   }$&$ 4.17  ^{+ 0.07  }_{ -0.08 }$&$ 0.11  ^{+ 0.49  }_{ -0.47 }$&$ 1.13  ^{+ 0.09  }_{ -0.11 }$&$ 1.44  ^{+ 0.16  }_{ -0.15 }$&$ 2.19  ^{+ 0.79  }_{ -0.60 }$&$ 0.53  \pm{ 0.18 }$ \\
K00041	  & 6521045    & 291.385986      & 41.990269       & 11.20 & 5909       & 4.30       & -0.12      &$ 5902  ^{+ 53    }_{ -74   }$&$ 3.62  ^{+ 0.20  }_{ -0.12 }$&$ -0.13 ^{+ 0.43  }_{ -0.38 }$&$ 1.23  ^{+ 0.21  }_{ -0.13 }$&$ 2.86  ^{+ 0.53  }_{ -0.65 }$&$ 4.73  ^{+ 3.63  }_{ -2.22 }$&$ 0.07  \pm{ 0.05 }$ \\
K00070	  & 6850504    & 287.697998      & 42.338718       & 12.50 & 5443       & 4.45       & -0.15      &$ 5444  ^{+ 44    }_{ -43   }$&$ 4.43  ^{+ 0.07  }_{ -0.07 }$&$ -0.17 ^{+ 0.52  }_{ -0.47 }$&$ 0.88  ^{+ 0.08  }_{ -0.12 }$&$ 0.94  ^{+ 0.09  }_{ -0.08 }$&$ 0.66  ^{+ 0.26  }_{ -0.19 }$&$ 1.49  \pm{ 0.45 }$ \\
K00072	  & 11904151   & 285.679382      & 50.241299       & 10.96 & 5627       & 4.39       & -0.81      &$ 5627  ^{+ 61    }_{ -53   }$&$ 4.35  ^{+ 0.06  }_{ -0.07 }$&$ -0.80 ^{+ 0.48  }_{ -0.51 }$&$ 0.79  ^{+ 0.11  }_{ -0.07 }$&$ 0.98  ^{+ 0.09  }_{ -0.08 }$&$ 0.67  ^{+ 0.28  }_{ -0.16 }$&$ 1.16  \pm{ 0.32 }$ \\
K00082	  & 10187017   & 281.482727      & 47.208031       & 11.49 & 4908       & 4.61       & 0.20       &$ 4909  ^{+ 41    }_{ -41   }$&$ 4.58  ^{+ 0.08  }_{ -0.07 }$&$ 0.18  ^{+ 0.42  }_{ -0.47 }$&$ 0.81  ^{+ 0.04  }_{ -0.09 }$&$ 0.76  ^{+ 0.07  }_{ -0.07 }$&$ 0.28  ^{+ 0.05  }_{ -0.06 }$&$ 2.61  \pm{ 0.75 }$ \\
K00085	  & 5866724    & 288.688690      & 41.151180       & 11.02 & 6172       & 4.36       & -0.05      &$ 6170  ^{+ 47    }_{ -48   }$&$ 4.33  ^{+ 0.08  }_{ -0.07 }$&$ -0.03 ^{+ 0.48  }_{ -0.48 }$&$ 1.18  ^{+ 0.12  }_{ -0.15 }$&$ 1.22  ^{+ 0.13  }_{ -0.11 }$&$ 1.76  ^{+ 0.72  }_{ -0.46 }$&$ 0.92  \pm{ 0.30 }$ \\
\enddata

\tablecomments{Columns marked with a $^{\dagger}$ are values from NASA Exoplanet Archive. Stars marked with a $^{\dagger\dagger}$ have missing $T_{\rm{eff}}$. Their $T_{\rm{eff}}$ values are estimated based on $g-r$ and $g-i$ values~\citep{Pinsonneault2012}, the adopted $T_{\rm{eff}}$ is the average of the results. There are three exceptions, values of $T_{\rm{eff}}$ for K03158 and K0368 are adopted from~\citet{Ammons2006} and $T_{\rm{eff}}$ for K04021 is based on the infrared flux method in ~\citet{Pinsonneault2012}}

\end{deluxetable}

\clearpage
 \end{landscape}
\newpage

\clearpage

\begin{landscape}
\begin{deluxetable}{lccccccccccccccccc}
\tablewidth{0pt}
\tabletypesize{\tiny}
\setlength{\tabcolsep}{0.02in}
\tablecaption{Orbital Parameters\label{tab:orbital_params}}
\tablehead{
\colhead{\textbf{KOI}} &
\colhead{\textbf{KIC}} &
\colhead{\textbf{$T_0$}} &
\colhead{\textbf{Period}} &
\colhead{\textbf{Impact}} &
\colhead{\textbf{R$_{\rm PL}$/R$_{\ast}$}} &
\colhead{\textbf{e}} &
\colhead{\textbf{$\omega$}} &
\colhead{\textbf{R$_{\rm PL}$}} &
\colhead{\textbf{Inclination}} &
\colhead{\textbf{a/R$_{\ast}$}} &
\colhead{\textbf{a}} &
\colhead{\textbf{$T_{\rm PL}$}} &
\colhead{\textbf{Depth}} &
\colhead{\textbf{Dur}} &
\colhead{\textbf{Sig}} &
\colhead{\textbf{p1}} &
\colhead{\textbf{p2}} \\

\colhead{\textbf{}} &
\colhead{\textbf{}} &
\colhead{\textbf{(BKJD)}} &
\colhead{\textbf{(d)}} &
\colhead{\textbf{Parameter}} &
\colhead{\textbf{}} &
\colhead{\textbf{}} &
\colhead{\textbf{(radian)}} &
\colhead{\textbf{(R$_{\oplus}$)}} &
\colhead{\textbf{($\deg$)}} &
\colhead{\textbf{}} &
\colhead{\textbf{(AU)}} &
\colhead{\textbf{(K)}} &
\colhead{\textbf{(ppm)}} &
\colhead{\textbf{(hr)}} &
\colhead{\textbf{($\sigma$)}} &
\colhead{\textbf{}} &
\colhead{\textbf{}} 

}

\startdata

\textbf{K00005.01}  & 8554498    & 132.9733  &$    4.78033 ^{+    0.00001 }_{   -0.00001 }$&$ 0.93  ^{+ 0.01  }_{ -0.01 }$&$ 0.045  ^{+ 0.001  }_{ -0.002 }$&$ 0.00  ^{+ 0.02  }_{ -0.00 }$&$ 0.00  ^{+ 0.13  }_{ -0.00 }$&$ 8.07  \pm{ 0.64  }$&$ 82.72 ^{+ 0.61  }_{ -0.46 }$&$ 7.6   ^{+ 0.5   }_{ -0.4  }$&$ 0.059  \pm{ 0.002  }$&$ 1434  \pm{ 45    }$&$ 983   \pm{ 3     }$& 2.3   &\textbf{ 5.1  }&0.932&\textbf{0.997} \\
K00005.02  & 8554498    & 133.3670  &$    7.05186 ^{+    0.00000 }_{   -0.00000 }$&$ 0.92  ^{+ 0.16  }_{ -0.17 }$&$ 0.003  ^{+ 0.002  }_{ -0.001 }$&$ 0.00  ^{+ 0.05  }_{ -0.00 }$&$ 0.04  ^{+ 0.21  }_{ -0.04 }$&$ 0.51  \pm{ 0.26  }$&$ 85.30 ^{+ 0.72  }_{ -0.60 }$&$ 11.2  ^{+ 1.0   }_{ -1.0  }$&$ 0.075  \pm{ 0.002  }$&$ 1159  \pm{ 69    }$&$ 12    \pm{ 4     }$& 1.9   & 0.4  &0.015&0.278\\
K00041.01  & 6521045    & 122.9506  &$   12.81574 ^{+    0.00004 }_{   -0.00005 }$&$ 0.57  ^{+ 0.20  }_{ -0.55 }$&$ 0.014  ^{+ 0.001  }_{ -0.001 }$&$ 0.00  ^{+ 0.30  }_{ -0.00 }$&$ 0.00  ^{+ 0.88  }_{ -0.00 }$&$ 3.08  \pm{ 0.74  }$&$ 87.33 ^{+ 2.57  }_{ -2.06 }$&$ 12.0  ^{+ 2.5   }_{ -2.4  }$&$ 0.114  \pm{ 0.005  }$&$ 1120  \pm{ 130   }$&$ 219   \pm{ 5     }$& 6.7   & 1.5  &0.615&0.988\\
K00041.02  & 6521045    & 133.1758  &$    6.88710 ^{+    0.00016 }_{   -0.00015 }$&$ 0.90  ^{+ 0.02  }_{ -0.09 }$&$ 0.009  ^{+ 0.001  }_{ -0.001 }$&$ 0.11  ^{+ 0.19  }_{ -0.11 }$&$ 0.01  ^{+ 1.11  }_{ -0.01 }$&$ 3.13  \pm{ 0.58  }$&$ 81.11 ^{+ 1.65  }_{ -1.50 }$&$ 5.4   ^{+ 1.3   }_{ -0.5  }$&$ 0.076  \pm{ 0.003  }$&$ 1628  \pm{ 119   }$&$ 72    \pm{ 4     }$& 4.4   & 0.3  &0.067&0.783\\
K00041.03  & 6521045    & 153.9839  &$   35.33314 ^{+    0.00109 }_{   -0.00204 }$&$ 0.93  ^{+ 0.10  }_{ -0.14 }$&$ 0.012  ^{+ 0.000  }_{ -0.003 }$&$ 0.00  ^{+ 0.22  }_{ -0.00 }$&$ 0.00  ^{+ 3.09  }_{ -0.00 }$&$ 3.78  \pm{ 0.88  }$&$ 86.88 ^{+ 0.83  }_{ -0.55 }$&$ 17.0  ^{+ 4.8   }_{ -2.4  }$&$ 0.226  \pm{ 0.010  }$&$ 947   \pm{ 106   }$&$ 96    \pm{ 10    }$& 6.3   & 1.0  &0.077&0.807\\
\enddata

\tablecomments{The last three columns are significance of pixel centroid offset between in- and out-of-transit, planet confidence before and after considering the planetary likelihood boost in~\citet{Lissauer2012}. Planet candidates with significant pixel centroid offset ($> 3\sigma$) and planet candidates with higher than 0.997 planet confidence are marked in bold.}

\end{deluxetable}

\clearpage
 \end{landscape}
\newpage

\clearpage

\begin{deluxetable}{lcccc}
\tablewidth{0pt}
\tablecaption{Visual companion detections with UKIRT images.\label{tab:AO_params}}
\tablehead{
\colhead{\textbf{KOI}} &
\colhead{\textbf{$\Delta$ K$_P$}} &
\colhead{\textbf{Separation}} &
\colhead{\textbf{Significance}} &
\colhead{\textbf{PA}} \\
\colhead{\textbf{}} &
\colhead{\textbf{(mag)}} &
\colhead{\textbf{(arcsec)}} &
\colhead{\textbf{($\sigma$)}} &
\colhead{\textbf{(deg)}} \\

}

\startdata

    K00005  &  ...  &  ...  &       ...  &       ... \\
    K00041  &  ...  &  ...  &       ...  &       ... \\
    K00070  &  4.7  &  7.7  &      57.7  &      51.4 \\
    K00072  &  ...  &  ...  &       ...  &       ... \\
    K00082  &  7.3  & 12.3  &       5.1  &     135.9 \\
    K00085  &  7.5  & 19.3  &       6.6  &     304.8 \\
    K00089  &  7.5  & 12.1  &       5.1  &      43.8 \\
    K00089  &  6.3  &  8.8  &      10.3  &     207.4 \\
\enddata

\end{deluxetable}

\clearpage

\newpage

\begin{deluxetable}{lccccc}
\tablewidth{0pt}
\tablecaption{RV measurement results for 23 KOIs.\label{tab:rv}}
\tablehead{
\colhead{{KOI}} &
\colhead{{$t_{\rm{start}}$}} &
\colhead{{$t_{\rm{end}}$}} &
\colhead{{$n_{\rm{obs}}$}} &
\colhead{{RMS}} &
\colhead{{$\delta v$}} \\
\colhead{{}} &
\colhead{{(MJD)}} &
\colhead{{(MJD)}} &
\colhead{{}} &
\colhead{{(m$\cdot$s$^{-1}$)}} &
\colhead{{(m$\cdot$s$^{-1}$)}} \\

}

\startdata


00005$^\dagger$&54983.515714&55843.325819&18&98.2&6.3 \\
00041&54988.510954&55814.257203&41&5.7&3.6 \\
00070&55073.385713&55723.555721&30&7.0&2.6 \\
00072&55074.377797&55414.302550&40&3.5&3.1 \\
00082&55311.578536&55814.275163&22&3.5&2.3 \\
00085&55696.490478&55738.515810&6&10.1&2.8 \\
00102&55107.034592&55381.183923&5&40.0&7.4 \\
00108&55134.287135&55809.400772&13&4.1&2.9 \\
\enddata

\tablecomments{$^\dagger$ marks the star with linear trend indicating a stellar companions. $\ast$ indicates the stars with a non-transiting planet companion~\citep[KOI 148, Howard Isaacson, private communication; KOI 246,][]{Gilliland2013}. }

\end{deluxetable}


\end{document}